\pdfoutput=1

\documentclass[11pt,twoside,a4paper,cmspaper,final,collab]{cms-tdr}
\def\svnVersion{22989:23024M}\def\cmsCernNo{1020-048}\def\cmsCernDate{\today}\def\cmsMessage{Submitted to the Journal of High Energy Physics (JHEP)}

\begin{document}\cmsNoteHeader{QCD-10-004}

\hyphenation{env-iron-men-tal}
\hyphenation{had-ron-i-za-tion}
\hyphenation{cal-or-i-me-ter}
\hyphenation{de-vices}
\RCS$Revision: 22995 $
\RCS$HeadURL: svn+ssh://alverson@svn.cern.ch/reps/tdr2/papers/QCD-10-004/trunk/QCD-10-004.tex $
\RCS$Id: QCD-10-004.tex 22995 2010-11-24 18:27:36Z rougny $
%
%
%

\providecommand {\etal}{\mbox{et al.}\xspace} 
\providecommand {\ie}{\mbox{i.e.}\xspace}     
\providecommand {\eg}{\mbox{e.g.}\xspace}     
\providecommand {\etc}{\mbox{etc.}\xspace}     
\providecommand {\vs}{\mbox{\sl vs.}\xspace}      
\providecommand {\mdash}{\ensuremath{\mathrm{-}}} 

\providecommand {\Lone}{Level-1\xspace} 
\providecommand {\Ltwo}{Level-2\xspace}
\providecommand {\Lthree}{Level-3\xspace}

\providecommand{\ACERMC} {\textsc{AcerMC}\xspace}
\providecommand{\ALPGEN} {{\textsc{alpgen}}\xspace}
\providecommand{\CHARYBDIS} {{\textsc{charybdis}}\xspace}
\providecommand{\CMKIN} {\textsc{cmkin}\xspace}
\providecommand{\CMSIM} {{\textsc{cmsim}}\xspace}
\providecommand{\CMSSW} {{\textsc{cmssw}}\xspace}
\providecommand{\COBRA} {{\textsc{cobra}}\xspace}
\providecommand{\COCOA} {{\textsc{cocoa}}\xspace}
\providecommand{\COMPHEP} {\textsc{CompHEP}\xspace}
\providecommand{\EVTGEN} {{\textsc{evtgen}}\xspace}
\providecommand{\FAMOS} {{\textsc{famos}}\xspace}
\providecommand{\GARCON} {\textsc{garcon}\xspace}
\providecommand{\GARFIELD} {{\textsc{garfield}}\xspace}
\providecommand{\GEANE} {{\textsc{geane}}\xspace}
\providecommand{\GEANTfour} {{\textsc{geant4}}\xspace}
\providecommand{\GEANTthree} {{\textsc{geant3}}\xspace}
\providecommand{\GEANT} {{\textsc{geant}}\xspace}
\providecommand{\HDECAY} {\textsc{hdecay}\xspace}
\providecommand{\HERWIG} {{\textsc{herwig}}\xspace}
\providecommand{\HIGLU} {{\textsc{higlu}}\xspace}
\providecommand{\HIJING} {{\textsc{hijing}}\xspace}
\providecommand{\IGUANA} {\textsc{iguana}\xspace}
\providecommand{\ISAJET} {{\textsc{isajet}}\xspace}
\providecommand{\ISAPYTHIA} {{\textsc{isapythia}}\xspace}
\providecommand{\ISASUGRA} {{\textsc{isasugra}}\xspace}
\providecommand{\ISASUSY} {{\textsc{isasusy}}\xspace}
\providecommand{\ISAWIG} {{\textsc{isawig}}\xspace}
\providecommand{\MADGRAPH} {\textsc{MadGraph}\xspace}
\providecommand{\MCATNLO} {\textsc{mc@nlo}\xspace}
\providecommand{\MCFM} {\textsc{mcfm}\xspace}
\providecommand{\MILLEPEDE} {{\textsc{millepede}}\xspace}
\providecommand{\ORCA} {{\textsc{orca}}\xspace}
\providecommand{\OSCAR} {{\textsc{oscar}}\xspace}
\providecommand{\PHOTOS} {\textsc{photos}\xspace}
\providecommand{\PROSPINO} {\textsc{prospino}\xspace}
\providecommand{\PYTHIA} {{\textsc{pythia}}\xspace}
\providecommand{\SHERPA} {{\textsc{sherpa}}\xspace}
\providecommand{\TAUOLA} {\textsc{tauola}\xspace}
\providecommand{\TOPREX} {\textsc{TopReX}\xspace}
\providecommand{\XDAQ} {{\textsc{xdaq}}\xspace}

\providecommand {\DZERO}{D\O\xspace}     


\providecommand{\de}{\ensuremath{^\circ}}
\providecommand{\ten}[1]{\ensuremath{\times \text{10}^\text{#1}}}
\providecommand{\unit}[1]{\ensuremath{\text{\,#1}}\xspace}
\providecommand{\mum}{\ensuremath{\,\mu\text{m}}\xspace}
\providecommand{\micron}{\ensuremath{\,\mu\text{m}}\xspace}
\providecommand{\cm}{\ensuremath{\,\text{cm}}\xspace}
\providecommand{\mm}{\ensuremath{\,\text{mm}}\xspace}
\providecommand{\mus}{\ensuremath{\,\mu\text{s}}\xspace}
\providecommand{\keV}{\ensuremath{\,\text{ke\hspace{-.08em}V}}\xspace}
\providecommand{\MeV}{\ensuremath{\,\text{Me\hspace{-.08em}V}}\xspace}
\providecommand{\GeV}{\ensuremath{\,\text{Ge\hspace{-.08em}V}}\xspace}
\providecommand{\TeV}{\ensuremath{\,\text{Te\hspace{-.08em}V}}\xspace}
\providecommand{\PeV}{\ensuremath{\,\text{Pe\hspace{-.08em}V}}\xspace}
\providecommand{\keVc}{\ensuremath{{\,\text{ke\hspace{-.08em}V\hspace{-0.16em}/\hspace{-0.08em}}c}}\xspace}
\providecommand{\MeVc}{\ensuremath{{\,\text{Me\hspace{-.08em}V\hspace{-0.16em}/\hspace{-0.08em}}c}}\xspace}
\providecommand{\GeVc}{\ensuremath{{\,\text{Ge\hspace{-.08em}V\hspace{-0.16em}/\hspace{-0.08em}}c}}\xspace}
\providecommand{\TeVc}{\ensuremath{{\,\text{Te\hspace{-.08em}V\hspace{-0.16em}/\hspace{-0.08em}}c}}\xspace}
\providecommand{\keVcc}{\ensuremath{{\,\text{ke\hspace{-.08em}V\hspace{-0.16em}/\hspace{-0.08em}}c^\text{2}}}\xspace}
\providecommand{\MeVcc}{\ensuremath{{\,\text{Me\hspace{-.08em}V\hspace{-0.16em}/\hspace{-0.08em}}c^\text{2}}}\xspace}
\providecommand{\GeVcc}{\ensuremath{{\,\text{Ge\hspace{-.08em}V\hspace{-0.16em}/\hspace{-0.08em}}c^\text{2}}}\xspace}
\providecommand{\TeVcc}{\ensuremath{{\,\text{Te\hspace{-.08em}V\hspace{-0.16em}/\hspace{-0.08em}}c^\text{2}}}\xspace}

\providecommand{\pbinv} {\mbox{\ensuremath{\,\text{pb}^\text{$-$1}}}\xspace}
\providecommand{\fbinv} {\mbox{\ensuremath{\,\text{fb}^\text{$-$1}}}\xspace}
\providecommand{\nbinv} {\mbox{\ensuremath{\,\text{nb}^\text{$-$1}}}\xspace}
\providecommand{\percms}{\ensuremath{\,\text{cm}^\text{$-$2}\,\text{s}^\text{$-$1}}\xspace}
\providecommand{\lumi}{\ensuremath{\mathcal{L}}\xspace}
\providecommand{\Lumi}{\ensuremath{\mathcal{L}}\xspace}
%
\providecommand{\LvLow}  {\ensuremath{\mathcal{L}=\text{10}^\text{32}\,\text{cm}^\text{$-$2}\,\text{s}^\text{$-$1}}\xspace}
\providecommand{\LLow}   {\ensuremath{\mathcal{L}=\text{10}^\text{33}\,\text{cm}^\text{$-$2}\,\text{s}^\text{$-$1}}\xspace}
\providecommand{\lowlumi}{\ensuremath{\mathcal{L}=\text{2}\times \text{10}^\text{33}\,\text{cm}^\text{$-$2}\,\text{s}^\text{$-$1}}\xspace}
\providecommand{\LMed}   {\ensuremath{\mathcal{L}=\text{2}\times \text{10}^\text{33}\,\text{cm}^\text{$-$2}\,\text{s}^\text{$-$1}}\xspace}
\providecommand{\LHigh}  {\ensuremath{\mathcal{L}=\text{10}^\text{34}\,\text{cm}^\text{$-$2}\,\text{s}^\text{$-$1}}\xspace}
\providecommand{\hilumi} {\ensuremath{\mathcal{L}=\text{10}^\text{34}\,\text{cm}^\text{$-$2}\,\text{s}^\text{$-$1}}\xspace}


\providecommand{\zp}{\ensuremath{\mathrm{Z}^\prime}\xspace}


\providecommand{\kt}{\ensuremath{k_{\mathrm{T}}}\xspace}
\providecommand{\BC}{\ensuremath{\mathrm{B_{c}}}\xspace}
\providecommand{\bbarc}{\ensuremath{\mathrm{\overline{b}c}}\xspace}
\providecommand{\bbbar}{\ensuremath{\mathrm{b\overline{b}}}\xspace}
\providecommand{\ccbar}{\ensuremath{\mathrm{c\overline{c}}}\xspace}
\providecommand{\JPsi}{\ensuremath{\mathrm{J}\hspace{-.08em}/\hspace{-.14em}\psi}\xspace}
\providecommand{\bspsiphi}{\ensuremath{\mathrm{B_s} \to \JPsi\, \phi}\xspace}
\providecommand{\AFB}{\ensuremath{A_\text{FB}}\xspace}
\providecommand{\EE}{\ensuremath{\mathrm{e^+e^-}}\xspace}
\providecommand{\MM}{\ensuremath{\mu^+\mu^-}\xspace}
\providecommand{\TT}{\ensuremath{\tau^+\tau^-}\xspace}
\providecommand{\wangle}{\ensuremath{\sin^{2}\theta_{\text{eff}}^\text{lept}(M^2_\mathrm{Z})}\xspace}
\providecommand{\ttbar}{\ensuremath{\mathrm{t\overline{t}}}\xspace}
\providecommand{\stat}{\ensuremath{\,\text{(stat.)}}\xspace}
\providecommand{\syst}{\ensuremath{\,\text{(syst.)}}\xspace}

\providecommand{\HGG}{\ensuremath{\mathrm{H}\to\gamma\gamma}}
\providecommand{\gev}{\GeV}
\providecommand{\GAMJET}{\ensuremath{\gamma + \text{jet}}}
\providecommand{\PPTOJETS}{\ensuremath{\mathrm{pp}\to\text{jets}}}
\providecommand{\PPTOGG}{\ensuremath{\mathrm{pp}\to\gamma\gamma}}
\providecommand{\PPTOGAMJET}{\ensuremath{\mathrm{pp}\to\gamma + \mathrm{jet}}}
\providecommand{\MH}{\ensuremath{M_{\mathrm{H}}}}
\providecommand{\RNINE}{\ensuremath{R_\mathrm{9}}}
\providecommand{\DR}{\ensuremath{\Delta R}}


\providecommand{\PT}{\ensuremath{p_{\mathrm{T}}}\xspace}
\providecommand{\pt}{\ensuremath{p_{\mathrm{T}}}\xspace}
\providecommand{\ET}{\ensuremath{E_{\mathrm{T}}}\xspace}
\providecommand{\HT}{\ensuremath{H_{\mathrm{T}}}\xspace}
\providecommand{\et}{\ensuremath{E_{\mathrm{T}}}\xspace}
\providecommand{\Em}{\ensuremath{E\hspace{-0.6em}/}\xspace}
\providecommand{\Pm}{\ensuremath{p\hspace{-0.5em}/}\xspace}
\providecommand{\PTm}{\ensuremath{{p}_\mathrm{T}\hspace{-1.02em}/}\xspace}
\providecommand{\PTslash}{\ensuremath{{p}_\mathrm{T}\hspace{-1.02em}/}\xspace}
\providecommand{\ETm}{\ensuremath{E_{\mathrm{T}}^{\text{miss}}}\xspace}
\providecommand{\ETslash}{\ensuremath{E_{\mathrm{T}}\hspace{-1.1em}/}\xspace}
\providecommand{\MET}{\ensuremath{E_{\mathrm{T}}^{\text{miss}}}\xspace}
\providecommand{\ETmiss}{\ensuremath{E_{\mathrm{T}}^{\text{miss}}}\xspace}
\providecommand{\VEtmiss}{\ensuremath{{\vec E}_{\mathrm{T}}^{\text{miss}}}\xspace}

\providecommand{\dd}[2]{\ensuremath{\frac{\mathrm{d} #1}{\mathrm{d} #2}}}

%

\providecommand{\ga}{\ensuremath{\gtrsim}}
\providecommand{\la}{\ensuremath{\lesssim}}
\providecommand{\swsq}{\ensuremath{\sin^2\theta_\mathrm{W}}\xspace}
\providecommand{\cwsq}{\ensuremath{\cos^2\theta_\mathrm{W}}\xspace}
\providecommand{\tanb}{\ensuremath{\tan\beta}\xspace}
\providecommand{\tanbsq}{\ensuremath{\tan^{2}\beta}\xspace}
\providecommand{\sidb}{\ensuremath{\sin 2\beta}\xspace}
\providecommand{\alpS}{\ensuremath{\alpha_S}\xspace}
\providecommand{\alpt}{\ensuremath{\tilde{\alpha}}\xspace}

\providecommand{\QL}{\ensuremath{\mathrm{Q}_\mathrm{L}}\xspace}
\providecommand{\sQ}{\ensuremath{\tilde{\mathrm{Q}}}\xspace}
\providecommand{\sQL}{\ensuremath{\tilde{\mathrm{Q}}_\mathrm{L}}\xspace}
\providecommand{\ULC}{\ensuremath{\mathrm{U}_\mathrm{L}^\mathrm{C}}\xspace}
\providecommand{\sUC}{\ensuremath{\tilde{\mathrm{U}}^\mathrm{C}}\xspace}
\providecommand{\sULC}{\ensuremath{\tilde{\mathrm{U}}_\mathrm{L}^\mathrm{C}}\xspace}
\providecommand{\DLC}{\ensuremath{\mathrm{D}_\mathrm{L}^\mathrm{C}}\xspace}
\providecommand{\sDC}{\ensuremath{\tilde{\mathrm{D}}^\mathrm{C}}\xspace}
\providecommand{\sDLC}{\ensuremath{\tilde{\mathrm{D}}_\mathrm{L}^\mathrm{C}}\xspace}
\providecommand{\LL}{\ensuremath{\mathrm{L}_\mathrm{L}}\xspace}
\providecommand{\sL}{\ensuremath{\tilde{\mathrm{L}}}\xspace}
\providecommand{\sLL}{\ensuremath{\tilde{\mathrm{L}}_\mathrm{L}}\xspace}
\providecommand{\ELC}{\ensuremath{\mathrm{E}_\mathrm{L}^\mathrm{C}}\xspace}
\providecommand{\sEC}{\ensuremath{\tilde{\mathrm{E}}^\mathrm{C}}\xspace}
\providecommand{\sELC}{\ensuremath{\tilde{\mathrm{E}}_\mathrm{L}^\mathrm{C}}\xspace}
\providecommand{\sEL}{\ensuremath{\tilde{\mathrm{E}}_\mathrm{L}}\xspace}
\providecommand{\sER}{\ensuremath{\tilde{\mathrm{E}}_\mathrm{R}}\xspace}
\providecommand{\sFer}{\ensuremath{\tilde{\mathrm{f}}}\xspace}
\providecommand{\sQua}{\ensuremath{\tilde{\mathrm{q}}}\xspace}
\providecommand{\sUp}{\ensuremath{\tilde{\mathrm{u}}}\xspace}
\providecommand{\suL}{\ensuremath{\tilde{\mathrm{u}}_\mathrm{L}}\xspace}
\providecommand{\suR}{\ensuremath{\tilde{\mathrm{u}}_\mathrm{R}}\xspace}
\providecommand{\sDw}{\ensuremath{\tilde{\mathrm{d}}}\xspace}
\providecommand{\sdL}{\ensuremath{\tilde{\mathrm{d}}_\mathrm{L}}\xspace}
\providecommand{\sdR}{\ensuremath{\tilde{\mathrm{d}}_\mathrm{R}}\xspace}
\providecommand{\sTop}{\ensuremath{\tilde{\mathrm{t}}}\xspace}
\providecommand{\stL}{\ensuremath{\tilde{\mathrm{t}}_\mathrm{L}}\xspace}
\providecommand{\stR}{\ensuremath{\tilde{\mathrm{t}}_\mathrm{R}}\xspace}
\providecommand{\stone}{\ensuremath{\tilde{\mathrm{t}}_1}\xspace}
\providecommand{\sttwo}{\ensuremath{\tilde{\mathrm{t}}_2}\xspace}
\providecommand{\sBot}{\ensuremath{\tilde{\mathrm{b}}}\xspace}
\providecommand{\sbL}{\ensuremath{\tilde{\mathrm{b}}_\mathrm{L}}\xspace}
\providecommand{\sbR}{\ensuremath{\tilde{\mathrm{b}}_\mathrm{R}}\xspace}
\providecommand{\sbone}{\ensuremath{\tilde{\mathrm{b}}_1}\xspace}
\providecommand{\sbtwo}{\ensuremath{\tilde{\mathrm{b}}_2}\xspace}
\providecommand{\sLep}{\ensuremath{\tilde{\mathrm{l}}}\xspace}
\providecommand{\sLepC}{\ensuremath{\tilde{\mathrm{l}}^\mathrm{C}}\xspace}
\providecommand{\sEl}{\ensuremath{\tilde{\mathrm{e}}}\xspace}
\providecommand{\sElC}{\ensuremath{\tilde{\mathrm{e}}^\mathrm{C}}\xspace}
\providecommand{\seL}{\ensuremath{\tilde{\mathrm{e}}_\mathrm{L}}\xspace}
\providecommand{\seR}{\ensuremath{\tilde{\mathrm{e}}_\mathrm{R}}\xspace}
\providecommand{\snL}{\ensuremath{\tilde{\nu}_L}\xspace}
\providecommand{\sMu}{\ensuremath{\tilde{\mu}}\xspace}
\providecommand{\sNu}{\ensuremath{\tilde{\nu}}\xspace}
\providecommand{\sTau}{\ensuremath{\tilde{\tau}}\xspace}
\providecommand{\Glu}{\ensuremath{\mathrm{g}}\xspace}
\providecommand{\sGlu}{\ensuremath{\tilde{\mathrm{g}}}\xspace}
\providecommand{\Wpm}{\ensuremath{\mathrm{W}^{\pm}}\xspace}
\providecommand{\sWpm}{\ensuremath{\tilde{\mathrm{W}}^{\pm}}\xspace}
\providecommand{\Wz}{\ensuremath{\mathrm{W}^{0}}\xspace}
\providecommand{\sWz}{\ensuremath{\tilde{\mathrm{W}}^{0}}\xspace}
\providecommand{\sWino}{\ensuremath{\tilde{\mathrm{W}}}\xspace}
\providecommand{\Bz}{\ensuremath{\mathrm{B}^{0}}\xspace}
\providecommand{\sBz}{\ensuremath{\tilde{\mathrm{B}}^{0}}\xspace}
\providecommand{\sBino}{\ensuremath{\tilde{\mathrm{B}}}\xspace}
\providecommand{\Zz}{\ensuremath{\mathrm{Z}^{0}}\xspace}
\providecommand{\sZino}{\ensuremath{\tilde{\mathrm{Z}}^{0}}\xspace}
\providecommand{\sGam}{\ensuremath{\tilde{\gamma}}\xspace}
\providecommand{\chiz}{\ensuremath{\tilde{\chi}^{0}}\xspace}
\providecommand{\chip}{\ensuremath{\tilde{\chi}^{+}}\xspace}
\providecommand{\chim}{\ensuremath{\tilde{\chi}^{-}}\xspace}
\providecommand{\chipm}{\ensuremath{\tilde{\chi}^{\pm}}\xspace}
\providecommand{\Hone}{\ensuremath{\mathrm{H}_\mathrm{d}}\xspace}
\providecommand{\sHone}{\ensuremath{\tilde{\mathrm{H}}_\mathrm{d}}\xspace}
\providecommand{\Htwo}{\ensuremath{\mathrm{H}_\mathrm{u}}\xspace}
\providecommand{\sHtwo}{\ensuremath{\tilde{\mathrm{H}}_\mathrm{u}}\xspace}
\providecommand{\sHig}{\ensuremath{\tilde{\mathrm{H}}}\xspace}
\providecommand{\sHa}{\ensuremath{\tilde{\mathrm{H}}_\mathrm{a}}\xspace}
\providecommand{\sHb}{\ensuremath{\tilde{\mathrm{H}}_\mathrm{b}}\xspace}
\providecommand{\sHpm}{\ensuremath{\tilde{\mathrm{H}}^{\pm}}\xspace}
\providecommand{\hz}{\ensuremath{\mathrm{h}^{0}}\xspace}
\providecommand{\Hz}{\ensuremath{\mathrm{H}^{0}}\xspace}
\providecommand{\Az}{\ensuremath{\mathrm{A}^{0}}\xspace}
\providecommand{\Hpm}{\ensuremath{\mathrm{H}^{\pm}}\xspace}
\providecommand{\sGra}{\ensuremath{\tilde{\mathrm{G}}}\xspace}
\providecommand{\mtil}{\ensuremath{\tilde{m}}\xspace}
\providecommand{\rpv}{\ensuremath{\rlap{\kern.2em/}R}\xspace}
\providecommand{\LLE}{\ensuremath{LL\bar{E}}\xspace}
\providecommand{\LQD}{\ensuremath{LQ\bar{D}}\xspace}
\providecommand{\UDD}{\ensuremath{\overline{UDD}}\xspace}
\providecommand{\Lam}{\ensuremath{\lambda}\xspace}
\providecommand{\Lamp}{\ensuremath{\lambda'}\xspace}
\providecommand{\Lampp}{\ensuremath{\lambda''}\xspace}
\providecommand{\spinbd}[2]{\ensuremath{\bar{#1}_{\dot{#2}}}\xspace}

\providecommand{\MD}{\ensuremath{{M_\mathrm{D}}}\xspace}
\providecommand{\Mpl}{\ensuremath{{M_\mathrm{Pl}}}\xspace}
\providecommand{\Rinv} {\ensuremath{{R}^{-1}}\xspace} 
\providecommand{\PHOJET} {{\textsc{phojet}}\xspace}
\providecommand{\PYTHIADsixT} {{\textsc{pythia d6t}}\xspace}
\providecommand{\PYTHIAeight} {{\textsc{pythia 8}}\xspace}
\providecommand{\PYTHIAV} {{\textsc{pythia 6}}\xspace}
\cmsNoteHeader{QCD-10-004} 
\title{Charged particle multiplicities in pp interactions at $\sqrt{s} = 0.9$, 2.36, and 7\TeV}

\date{\today}

\abstract{
Measurements of primary charged hadron multiplicity distributions are presented
for non-single-diffractive events in proton-proton collisions at centre-of-mass energies of $\sqrt{s} = 0.9$, 2.36, and 7\TeV,
in five pseudorapidity ranges from $|\eta|<0.5$ to $|\eta|<2.4$.
The data were collected with the minimum-bias trigger of the CMS experiment during the LHC commissioning
runs in 2009 and the 7\TeV run in 2010. 
The multiplicity distribution at $\sqrt{s} = 0.9\TeV$ is in
agreement with previous measurements. 
At higher energies the increase of the mean multiplicity with 
$\sqrt{s}$ is underestimated by most event generators.
The average transverse momentum as a function of the multiplicity is also presented. 
The measurement of higher-order moments of the multiplicity
distribution confirms the violation of Koba-Nielsen-Olesen scaling that has been observed at lower energies.
}

\hypersetup{%
pdfauthor={CMS Collaboration},%
pdftitle={Charged particle multiplicities in pp interactions at $sqrt(s) = 0.9$, 2.36, and 7\TeV},%
pdfsubject={CMS, Charged particle multiplicity},%
pdfkeywords={CMS, physics, QCD, multiplicity, moments}}

\maketitle 

\section{Introduction}
\label{intro}

The charged hadron multiplicity, or number of primary charged hadrons,
$n$, is a basic global observable characterising final states in
high-energy-collision processes. The multiplicity distribution, $P_n$,
is the probability to produce $n$ charged hadrons in an event,
either in full phase space or in restricted phase space domains.
In this paper we report measurements of $P_n$ in non-single-diffractive~\cite{Kittel:2005fu}
proton-proton collisions, at centre-of-mass energies
$\sqrt{s} = 0.9$, 2.36, and 7\TeV at the Large Hadron Collider (LHC)~\cite{Evans:2008zzb}.
The measurements are based on events recorded by
the Compact Muon Solenoid (CMS)~\cite{:2008zzk} experiment, using a minimum-bias trigger.

Energy-momentum and charge conservation significantly influence the
multiplicity distribution for the full phase space. The distribution in
restricted phase space, which is less affected by such constraints,
is expected to be a more sensitive probe of the underlying
dynamics and can be used to better constrain phenomenological models.
Comprehensive reviews on the subject can be found in~\cite{Wolf:1995pc,Kittel:2005fu,GrosseOetringhaus:2009kz}.
The measurements described in this paper are performed for intervals of increasing
extent in pseudorapidity from $|\eta|<0.5$ up to $|\eta|<2.4$,
where $\eta$ is defined as $-$ln[tan($\theta$/2)],
and $\theta$ is the polar angle of the particle with respect to the
counterclockwise beam direction.
In these measurements primary charged hadrons are defined as all charged hadrons
produced in the interaction, including the products of strong and
electromagnetic decays, but excluding products of weak decays and
hadrons originating from secondary interactions.

Independent emission of single particles yields a Poissonian $P_n$.
Deviations from this shape, therefore, reveal
correlations. These correlations are predominantly short range in
rapidity, attributed to cluster decays, and reflect local conservation of
quantum numbers in the hadronisation process.
In hadron-hadron interactions, additional large long-range rapidity
correlations are observed, whose magnitude increases with $\sqrt{s}$~\cite{Capella:1978rg}.
In contrast, in e$^+$e$^-$ annihilations such long-range correlations are much weaker
and practically absent in two- and three-jet event
samples~\cite{Akers:1993tg}.

The mean of the multiplicity distribution, $\langle n\rangle $, is
equal to the integral of the inclusive single-particle density in
the considered phase-space domain. Higher-order moments of $P_n$ 
measure event-to-event multiplicity fluctuations. They are related to
the two-particle and higher-order inclusive density correlations and provide more
detailed dynamical information than that contained in single-particle
inclusive
spectra~\cite{Mueller:1971ez,Brown:1972bw,Biebl:1972kd,Webber:1972gc,Koba:1973zz}.
The average transverse momentum of the charged particles,  $\langle \pt\rangle$, exhibits a
positive correlation with the event multiplicity in hadron-hadron collisions
(\cite{modelliptm1,modelliptm2,modelliptm3,modelliptm4,modelliptm5,modelliptm6,Aaltonen:2009ne} and references therein).

Traditionally, the $s$ dependence of $P_n$ and
its moments has been much
discussed~\cite{Wolf:1995pc,Kittel:2005fu,GrosseOetringhaus:2009kz}
in relation to Koba-Nielsen-Olesen (KNO) scaling~\cite{Koba:1972ng, Hegyi:2000sp}.
In this framework, one studies the KNO function $\Psi(z)=\langle
n\rangle P_n$, where $z=n/\langle n\rangle $.
If KNO scaling holds, $\Psi(z)$ and the normalised
moments $C_q=\langle n^q\rangle / \langle n\rangle^q $ are
independent of ${s}$.

Throughout this paper, we compare the data with existing measurements at
similar or lower centre-of-mass energies and with predictions of the multiplicity distribution and its mean value from 
analytical and event generator Monte Carlo (MC) models.
The models are based on the assumption that hadrons are produced via the fragmentation of colour strings.
This comparison should allow for a better tuning of the existing MC models to
accurately simulate minimum-bias events and underlying-event effects.

The next section gives a short description of the CMS detector. Section 3 describes the MC models used in the analysis, while
Section 4 presents the data samples. The track reconstruction and acceptance are explained in Section 5. Section 6
describes the corrections applied to the data. Section 7 lists all relevant systematic
uncertainties. The results are discussed in Section 8.

\section{The CMS detector}
A complete description of the CMS detector can be found in~\cite{:2008zzk}. The
CMS experiment uses a right-handed coordinate system, with the origin at the
nominal interaction point (IP), the $x$ axis pointing to the centre of the LHC
ring, the $y$ axis pointing up, and the $z$ axis along the
counterclockwise beam direction.

The central feature of the CMS detector is a superconducting solenoid of 6 m
internal diameter, providing an axial magnetic field with nominal
strength of 3.8~T.
Immersed in the magnetic field are the pixel tracker, the silicon-strip
tracker (SST), the lead tungstate electromagnetic calorimeter, the
brass/scintillator hadron calorimeter and the muon detection system. In addition to the barrel and endcap
calorimeters, the steel/quartz-fibre forward calorimeter (HF) covers the region $2.9 < |\eta | < 5.2$.

Two of the CMS subdetectors acting as LHC beam monitors, the Beam Scintillation
Counters (BSC) and the Beam Pick-up Timing for the eXperiments (BPTX)
devices, were used to trigger the detector readout. The BSCs are located
along the beam line on each side of the IP at a distance of 10.86 m and are
sensitive in the range $3.23 < |\eta | < 4.65$. The two BPTX devices,
which are located inside the beam pipe at distances of 175 m from the IP, 
are designed to provide precise information on the bunch structure and
timing of the incoming beams, with a time resolution better than  0.2 ns.

The tracking detector consists of 1440 silicon-pixel and 15148
silicon-strip detector modules. The barrel part consists of 3 (10) layers of pixel (SST) modules around the
IP at distances ranging from 4.4 cm to 1.1 m. Five out of the
10 strip layers are double sided and provide additional $z$ coordinate measurements.
The two endcaps consist of 2 (12) disks of pixel (SST) modules that extend the 
pseudorapidity acceptance to $|\eta | < 2.5$. The
tracker provides an impact parameter resolution of about
$100\micron$ and a \pt resolution of about 0.7\% for 1\GeVc 
charged particles at normal incidence~\cite{Adam:2006ki, CMSTrackingperf}.

\section{Models}
The \PYTHIAV~\cite{Sjostrand:2006za} generator and its fragmentation model tuned to CDF
data~\cite{Affolder:2001xt,Bartalini:2008zz}, hereafter called \PYTHIADsixT,
is used as a baseline model to simulate inelastic pp collisions.
However, at $7\TeV$ a dedicated \PYTHIA tune~\cite{Moraes:2007rq} describing
better the high multiplicities is used for correcting the data.
Alternative tunings that differ mainly in the modelling of multiple
parton interactions have also been considered~\cite{Bartalini:2008zz, Buckley:2009bj,Skands:2010ak}.
\PHOJET~\cite{Engel:1994vs, Engel:1995yda} is used as an alternative event generator that differs mainly in the
underlying dynamical model for particle production. While \PYTHIA contains at least one hard scatter per event,
particle production in \PHOJET, which is based on the dual-parton model~\cite{Capella:1992yb}, is predominantly
soft and contains in general multiple-string configurations derived from the dual-parton
model with multi-Pomeron exchanges~\cite{Capella:1992yb}. Each Pomeron exchange
gives rise to two strings stretched between either valence
or sea partons. At low energies the dominant
process is single-Pomeron exchange, which leads
to two strings stretched between valence quarks and diquarks.
With increasing energy, additional Pomeron exchanges occur,
forming strings stretched between sea
partons. Because the sea partons carry on average only a
small fraction of the momentum of the incident hadrons,
these strings are concentrated in the central rapidity
region. These extra strings~\cite{PhysRevD.69.034007,Matinyan:1998ja} are needed to explain the
KNO scaling violations observed at high energies~\cite{Alner:1984is,Alner:1985wj}, the increase of the
central particle density with increasing energy~\cite{Alner:1984is,Alner:1985wj,Albajar:1989an}, the $\langle p_T\rangle$
versus $n$ dependence~\cite{Alexopoulos:1988na,PhysRevD.29.1007}, and long-range rapidity correlations~\cite{Ansorge:1988fg}.
The distribution of the number of exchanged Pomerons can be obtained
from perturbative Reggeon calculus and unitarity, by means of fits to the measured total,
elastic, and diffractive cross sections as described in~\cite{Gotsman:2009xm, Likhoded:2010pc, Kaidalov:2009hn}.
Other alternatives based on the Colour Glass Condensate (CGC) picture with saturated
gluons~\cite{Levin:2010dw} also make predictions for the multiplicity dependence of
$\langle \pt\rangle$ and for the long-range rapidity correlations.

We also compare our measurements with a new fragmentation model implemented in \PYTHIAeight~\cite{Sjostrand:2007gs}.
In addition to having \pt-ordered parton showers,
rather than an ordering by virtuality,  it differs from its predecessor in the modelling of multiple-parton interactions
and the treatment of beam remnants and diffraction.
The detailed MC simulation of the CMS detector response is based on
\GEANTfour~\cite{Agostinelli:2002hh}. Simulated events were processed and reconstructed in the
same manner as collision data.

\section{Data sample}
\label{selection}
Near the end of 2009, the CMS experiment collected two datasets of
proton-proton collisions at centre-of-mass energies of 0.9 and
2.36\TeV. In March 2010 a new running period at $\sqrt{s} = 7\TeV$ started,
of which data collected in the first days have been analysed for this paper.
The corresponding inelastic proton-proton interaction rates were about 11, 3, and 50 Hz,
respectively, for these datasets. 
At these rates, the fraction of bunch crossings in which 
two or more minimum-bias collisions occurred is negligible~\cite{Khachatryan:2010xs,PhysRevLett.105.022002}.

Diffraction is commonly characterised by one (single-diffraction) or two (double-diffraction)
colourless exchanges resulting in the observation of a large rapidity interval devoid of any hadron activity (rapidity gap).
All results presented in this paper refer to inelastic
non-single-diffractive (NSD) interactions and are based on an event
selection that retains a large fraction of the non-diffractive (ND) and
double-diffractive (DD) events, while disfavouring single-diffractive (SD) events.

The trigger and offline event selection are nearly identical to those used in~\cite{Khachatryan:2010xs,PhysRevLett.105.022002};
about three times more events were used in this analysis at
$\sqrt{s} = 0.9$ and 7\TeV.  
The trigger required a signal
in any of the BSC scintillator counters, in coincidence with either of the two BPTX devices 
indicating the presence of at least one proton bunch crossing the IP. 
Beam halo backgrounds are reduced by using the timing information from the
BSC counters at opposite ends of the CMS detector. Additional beam-induced backgrounds are removed by requiring the cluster sizes
in the pixel detector to be consistent with a single primary vertex, as 
described in~\cite{Khachatryan:2010xs}.
NSD events are selected by requiring at least one HF calorimeter tower with 
more than 3 GeV of total energy in each of the positive-$z$ and negative-$z$ HF calorimeters.
Finally, a reconstructed primary vertex is required with the $z$ coordinate within $\pm15$ cm of the centre of the beam collision region.

For the final analysis, totals of about 132, 12, and 442 thousand events  were retained at
$\sqrt{s} = 0.9$, 2.36, and 7\TeV, respectively. Event yields at various
stages of the selection are shown in
Table~\ref{tab1}.

\section{Track reconstruction and acceptance definition}
\label{trackingsection}
The barrel and endcap pixel and SST detectors are used in the reconstruction of tracks
within an acceptance of $|\eta|<2.5$.
Due to a large drop in reconstruction efficiency near the limits of this range, we restrict the computation of the multiplicity spectra to the region $|\eta | < 2.4$. 

In proton-proton collisions at the LHC, the events selected by minimum-bias
triggers involve predominantly soft interactions and contain mostly particles
with small transverse momenta. These are reconstructed by extending the
standard tracking algorithms of the CMS experiment,  which are based on a
combinatorial track finder~\cite{Adam:2006ki} that performs multiple
iterations. Hits that can be assigned unambiguously to tracks in one iteration are removed from
the collection of tracker hits to create a smaller collection that can be used in the subsequent
iteration. This iterative procedure was further optimised for primary track
reconstruction with $\pt \geq 100$~\MeVc in minimum-bias
events~\cite{Khachatryan:2010xs}. 
The reconstruction efficiency,
estimated by means of the detector simulation,  exceeds 90\% for tracks with
$\pt>500\MeVc$ and drops below 70\% for tracks
with $\pt<100\MeVc$. Contamination from mis-reconstructed tracks is below 5\% for $\pt<500\MeVc$. 
 
\begin{table}[t!]
\begin{center}
\caption{Event yields in each data sample after sequential trigger and event selection.\label{tab1}}
\begin{tabular}{|c|c|c|c|}
\multicolumn{4}{c}{}\\
\hline
\multirow{2}{*}{Selection} & \multicolumn{3}{c|}{$\sqrt{s}$~(TeV)} \\
\cline{2-4}
	& 0.9 & 2.36 & 7 \\\hline
\multicolumn{1}{|l|}{beam background rejection + L1 trigger}     & 254666 & 18739 & 610549 \\   
\multicolumn{1}{|l|}{all preceding + forward calorimeters}    & 146658 & 12019 & 500077 \\
\multicolumn{1}{|l|}{all preceding + primary vertex}		& 132294 & 11674 & 441924 \\     
\hline
\end{tabular}
\end{center}
\end{table}

After three iterations of the combinatorial track finder, the position 
of the primary vertex is recomputed and then used as an additional
constraint in a refit of all previously reconstructed tracks, thus 
improving the overall resolution in $\eta$ and \pt.
An agglomerative clustering algorithm followed by a Gaussian mixture
model~\cite{ferencvertex} were applied in order to optimise the
vertex-finding efficiency.

The contamination due to decays of long-lived particles (denoted as V$^{0}$ decays) is dominated
by charged pions and protons originating from
K$^0_S$ and $\Lambda$ decays. The K$^0_S$ production rate is roughly 5\% of that for all charged
particles~\cite{Becattini:2009ee}, and the resulting charged secondaries amount to 7\% of all $\pi^{\pm}$.
The $\Lambda$ production is
measured to be 43\% of K$^0_S$~\cite{Alpgard:1982kx, Alpgard:1982zz} in this kinematic domain,
yielding another 3\% of charged secondaries consisting of protons and
pions. In order to reduce the contamination of these V$^{0}$ decay
products and secondaries produced in interactions of charged particles
with the material of the detector, we require all reconstructed tracks to be associated with the primary
vertex.  This is done by selecting tracks with a small impact
parameter with respect to the primary vertex position both in the transverse
plane and along the $z$ axis~\cite{Khachatryan:2010xs}. Only 1.5\% of K$^0_S$ and $\Lambda$ decay products pass these
selections, resulting in a final contamination of 0.2\%. 

The K$^{\pm}/\pi^{\pm}$ ratio is known to be fairly constant over
a wide range of centre-of-mass energies and is between 8 and 12\%~\cite{Becattini:2009ee}. The K$^{\pm}$ have a lower reconstruction
efficiency at low \pt than pions and mismodeling of the K$^{\pm}/\pi^{\pm}$ ratio could
result in a change in the multiplicity distribution. However a doubling of the
K$^{\pm}/\pi^{\pm}$ ratio yields a negligible shift of 0.25\% in the multiplicity average.
This is expected because the $\langle \pt\rangle$ of K$^{\pm}$ is
substantially higher than $\pi^{\pm}$, and K$^{\pm}$ therefore contribute very 
little in the \pt range where the reconstruction efficiencies differ.
Finally, only tracks with a relative uncertainty on their measured transverse momentum smaller than 10\% are selected;
this requirement rejects mainly low-quality and badly reconstructed tracks.

\section{Corrections}
\label{corrections}
Due to the requirement of significant activity in both ends of the HF, the event selection acceptance for SD events is small: 5\% at 0.9\TeV and 7\% at 7\TeV.
This contribution is therefore
subtracted based on simulated \PYTHIA events. 
The \PYTHIA and \PHOJET predictions of the initial fractions of
SD events differ substantially, as seen from
Table~\ref{tab2}. The difference of the two predictions is taken as the systematic uncertainty related to the SD subtraction.

\begin{table}[t!]
\begin{center}
\caption{The percentage of single-diffractive events (SD/inelastic) at each centre-of-mass
energy as predicted by \PYTHIADsixT and \PHOJET.  \label{tab2}}
\hspace{0.5cm}
\begin{tabular}{|c|c|c|c|}
\multicolumn{4}{c}{}\\
\hline
\multirow{2}{*}{Generator} & \multicolumn{3}{c|}{$\sqrt{s}$~(TeV)} \\
\cline{2-4}
	& 0.9 & 2.36 & 7 \\\hline
\multicolumn{1}{|l|}{\PYTHIADsixT}	& 22.5 & 21.0 & 19.2 \\
\multicolumn{1}{|l|}{\PHOJET}    	& 18.9 & 16.2 & 13.8 \\   
\hline
\end{tabular}
\end{center}
\end{table}

It is customary to normalise the charged hadron 
multiplicity distribution $P_n=\sigma_n/\sigma$, where $\sigma_n$ denotes the cross section for a fixed
multiplicity $n$, to either the total 
inelastic cross section or the NSD cross section. For the 
results presented in this paper the normalisation factor 
$\sigma$ corresponds to the latter.

The minimum-bias trigger and NSD selection unavoidably introduce a bias in the measured charged hadron multiplicity.
Furthermore, a fraction of the events are removed by the
requirement of a good quality primary vertex.
These effects result in an accepted  multiplicity given by                                       
\begin{equation}
T_n=\epsilon_n \cdot P_n,
\end{equation}
where 
$\epsilon_n$ is the trigger and event reconstruction efficiency for multiplicity
$n$. This efficiency is close to 100\% for multiplicities larger
than $n=20$ and drops gradually to 40\% for $n=1$.
 
Due to inefficiencies in track reconstruction and acceptance, 
the creation of secondary particles by the interaction of primaries with the
beam pipe and the detector material, and the presence of decay products of
long-lived hadrons, one will 
in general not measure the true multiplicity $n$ but a statistically
related quantity $m$. 
The statistical distribution $O_m$ of this observed multiplicity is related to the true accepted multiplicity distribution by the linear relation
\begin{equation}
O_m=\sum_n R_{m,n} \cdot T_n.
\end{equation}
The problem of inverting the response matrix $R_{m,n}$, here taken from MC, is well known and extensive literature on
the topic exists~\cite{cowan:1998,bohm:2010}. When dealing with limited event samples, an algebraic inversion
of $R_{m,n}$ turns out to be impossible and leads to unstable results. We therefore use a
Bayesian unfolding method, as described in~\cite{D'Agostini:1994zf}. 
The unfolding procedure introduces large statistical correlations between
adjacent bins of the multiplicity spectrum. The full covariance matrix of the
unfolded multiplicity spectrum was calculated using a resampling technique~\cite{bohm:2010}. 

The average transverse momentum $\langle \pt\rangle $ as a function of $n$ was
measured as well. For each bin in raw multiplicity, a Monte Carlo based
correction factor $\langle \pt^{\mathrm{gen}}\rangle /\langle \pt^{\mathrm{rec}}\rangle $  was applied to convert the
measured $\langle \pt\rangle $ to the corresponding value for primary charged hadrons.
Subsequently, the response matrix $R_{m,n}$, weighted by the corrected data, was applied to correct the
multiplicities.

The correctness of the unfolding procedure was verified using MC events
generated with our baseline \PYTHIADsixT tune where
we compared the multiplicity distribution of generated primary hadrons with
that of reconstructed tracks after unfolding. At all energies considered, an excellent agreement with the generated hadrons is
achieved, proving the stability of the procedure. 

\begin{figure}[hbt!]
\begin{center}
\includegraphics[width=0.55\textwidth]{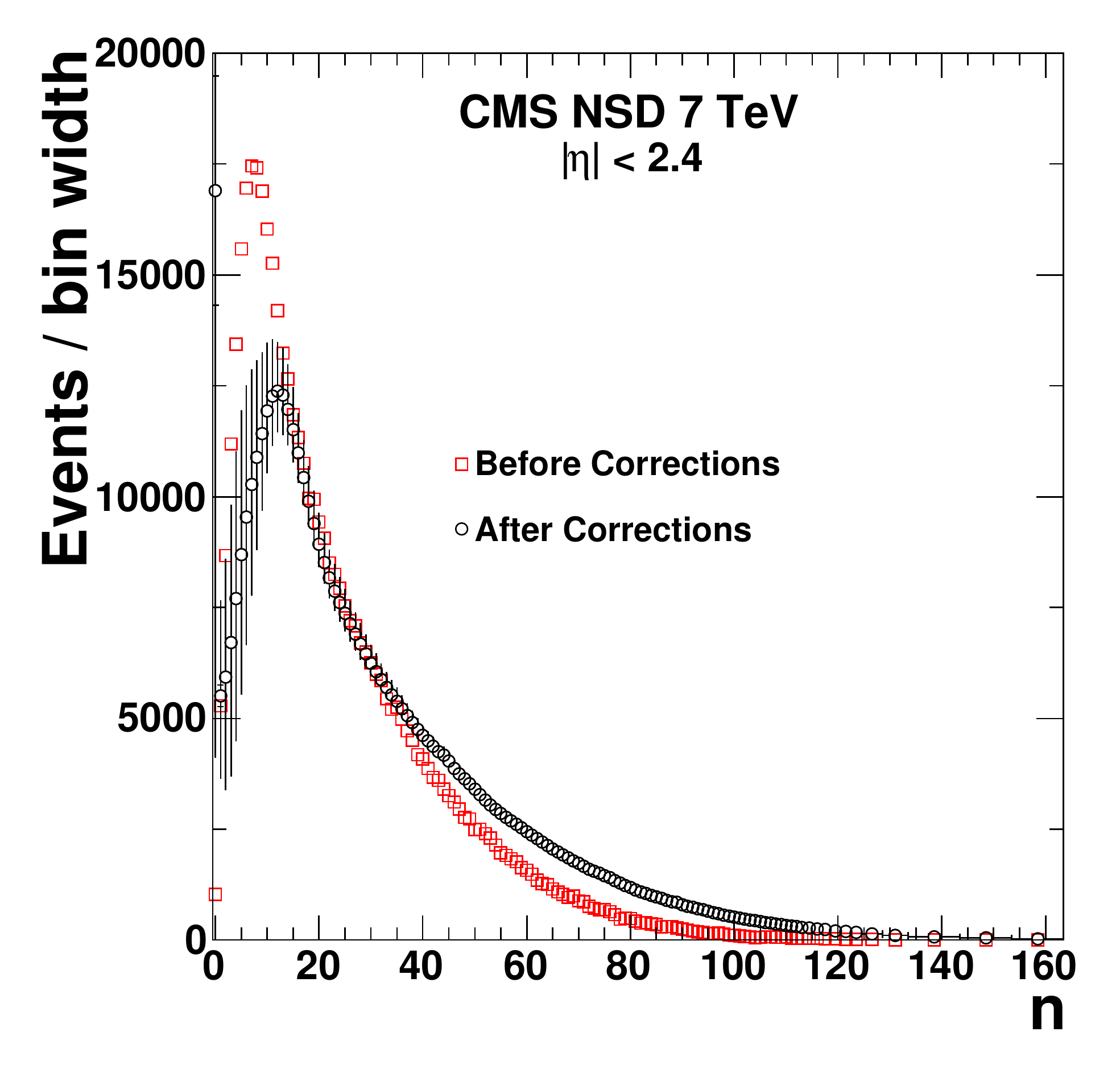}
\caption{A comparison of the uncorrected and fully corrected multiplicity
distribution at $\sqrt{s} = 7\TeV$ for $|\eta|<2.4$ . The uncertainties before corrections are
statistical only, while after corrections the statistical and systematic
uncertainties are added in quadrature. \label{showcorrection}}
\end{center}
\end{figure}

The track reconstruction efficiency of the  minimum-bias tracking 
drops drastically for $\pt<100$\MeVc, while the mis-reconstructed track rate increases. 
Rather than correcting for these effects using MC
simulations, we reconstruct the \pt spectrum in data themselves and
calculate the fraction of charged hadrons with
$\pt<100\MeVc$ by extrapolating the measured spectrum in data,
using a parametrisation based on an exponential of a third-degree polynomial in \pt.
The fraction of charged hadrons
added by this correction ranges between 5\% and  7\%
depending on the centre-of-mass energy and the pseudorapidity interval under study.
The functional form used to extrapolate to the lowest transverse momenta
introduces an uncertainty of 1\% on this fraction and is  taken into account in the systematic uncertainties.
The effect of all the
correction procedures on the measured raw multiplicity distribution is illustrated in 
Fig.~\ref{showcorrection} for a centre-of-mass energy of 7\TeV.

\section{Systematic uncertainties}\label{systematics}
Four main sources of systematic uncertainty contribute to the total
uncertainty on the multiplicity distribution $P_n$ and $\langle\pt\rangle$ versus $n$: the uncertainty on the
trigger and event selection efficiency, the  uncertainty on
the tracking efficiency and acceptance, the model dependence of the response
matrix, and the model dependent SD subtraction. All four are discussed below. The effects of misalignment of the
tracking detector, dead sensors, and the uncertainty on the vertex position
are much smaller and are contained within the overall tracking systematic
uncertainty.
For $\langle\pt\rangle$ versus $n$, the effect of the event selection and SD subtraction was observed to
be negligible.

{\bf Trigger and event selection efficiency.}
The corrections for the trigger and event selection efficiency are
based entirely on MC simulation. The largest impact on the overall 
efficiency is that of the HF coincidence requirement.
A cross-check of the multiplicity dependent
efficiency with zero-bias events, which are by definition not biased at all (random trigger on collisions), shows good agreement within statistical errors between data
and MC.
A relative shift of the efficiency correction factors by ${}^{+5\%}_{-7\%}$ for low multiplicities,
decreasing to $\pm1\%$ for $n\sim20$,
covers the trigger efficiency measured
in zero-bias data.
This leads to a maximum $5\%$ systematic uncertainty.

{\bf Tracking efficiency and acceptance.}
A correct description of the tracking efficiency in the MC
simulation of the detector is essential for obtaining a correct response
matrix. At low transverse momenta, the efficiency drops due to a loss of hits
on the tracks that are stopped within the tracking volume. As in~\cite{Khachatryan:2010xs}, we assign a 2\% uncertainty on this efficiency.
The remaining contamination is mostly due to secondary tracks originating from interactions with the
material of the LHC beam pipe around the interaction point. This is
estimated in~\cite{Khachatryan:2010xs} to be no more than 1\%, and is confirmed based on~\cite{TRK-10-003}. 
These uncertainties, together with smaller contributions from misalignment of the tracking detector,
beam-halo background, multiple counting of tracks, and mis-reconstructed tracks
are added in quadrature to produce a total tracking uncertainty of 2.5\%.
All correction factors related to track reconstruction are summarised in
Table~\ref{systable}.
As was discussed in Section~\ref{trackingsection}, the contamination from V$^0$ decays
after associating tracks with the primary vertex is small (0.2\%) and already included
in the systematic uncertainty for secondary tracks. The
difference in reconstruction efficiency for charged kaons and pions also has
a negligible impact on the measured multiplicity distribution.
Finally, the extrapolation uncertainty from $\pt = 100\MeVc$ to zero is 1\% (Section~\ref{corrections}).
For $\langle \pt\rangle $ versus $n$, this was taken into account by changing
the mean transverse momentum by the average \pt of tracks lost during the event reconstruction folded with the
tracking efficiency uncertainty. The overall effect of this systematic uncertainty is less than $10\%$ for small multiplicities,
but can reach up to $30\%$ for
the high multiplicities, where it is the main source of systematic uncertainty.

\begin{table}[t!]
\begin{center}
\caption{\label{systable}Summary of systematic uncertainties on the track reconstruction.}
\begin{tabular}{|c|c|}
\multicolumn{2}{c}{}\\
\hline
 Source & tracking uncertainty (\%)    \\
\hline
 Tracking efficiency
         & 2.0 \\
Acceptance & 1.0 \\
Pixel hit efficiency & 0.3 \\
Pixel cluster splitting & 0.2 \\
Correction for secondaries & 1.0  \\
Misalignment & 0.1 \\
Beam halo & 0.1 \\
Multiple track counting & 0.1 \\
Mis-reconstructed tracks & 0.5 \\
\hline
\hline
 Total  & 2.5 \\
\hline
\end{tabular}
\end{center}
 \end{table}

{\bf Model dependence.}
The baseline MC model that is used to unfold the multiplicity distribution
underestimates the single-particle densities at zero rapidity by 10\% relatively to $|\eta|\sim 1.5$ at $\sqrt{s} = 0.9$ and 2.36\TeV, but correctly
describes the 7\TeV data shapes. This discrepancy affects the
response matrix used in the unfolding procedure. The effect on the
multiplicity distribution is estimated to be at most 3\% by means of an alternative MC tune.
The robustness of the unfolding procedure was verified by unfolding
pseudo-data generated with \PHOJET using a response matrix constructed with
\PYTHIA. The induced variation on $P_n$ is below 3\% and is therefore
considered to be covered already by the systematic uncertainty related to modelling of the single-particle densities.

{\bf SD subtraction.}
Finally, the modelling of SD events was found to be different 
in \PYTHIA and \PHOJET, both in the predicted relative fraction and in the predicted
multiplicity distribution. The largest variation is obtained by
subtracting the \PHOJET multiplicity distribution for SD events, using the
event fraction predicted by \PYTHIA and vice versa. The impact can be as
large as 20\% in some low-multiplicity bins. At high multiplicities this uncertainty
is below 5\%.

All systematic uncertainties are calculated in each multiplicity bin for upward and downward
effects on the multiplicity distribution and added in quadrature
to yield a total systematic uncertainty. The total uncertainty on
$P_n$ is below 10\% for a large part of
the multiplicity distribution, but increases to 40\% at the lowest and highest
multiplicities.

\section{Results}\label{results}
\subsection{Charged hadron multiplicity distributions}\label{result1}
\begin{figure}[hbt!]
\begin{center}
\begin{tabular}{cc}
\includegraphics[width=0.49\textwidth]{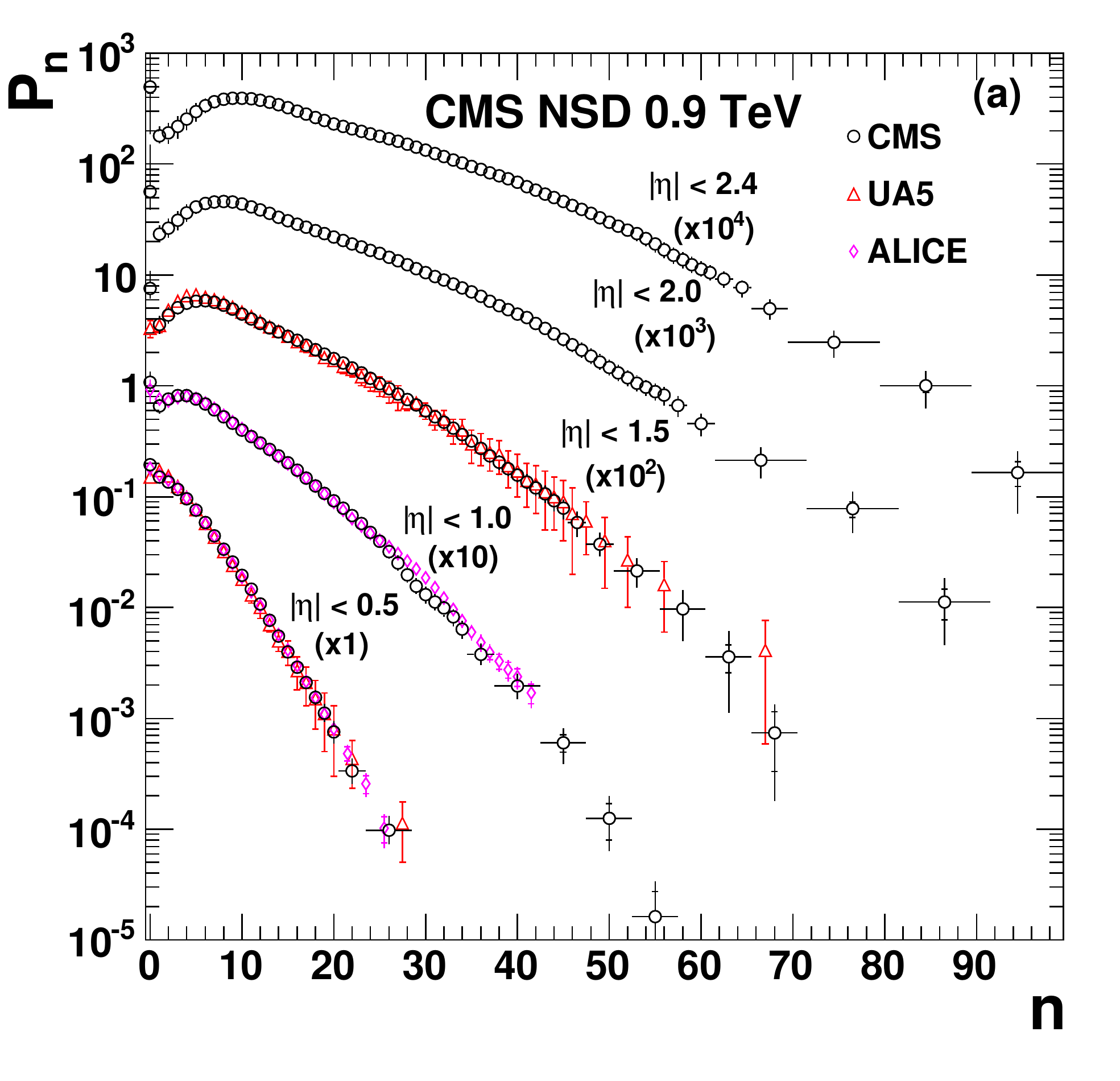} &
\includegraphics[width=0.49\textwidth]{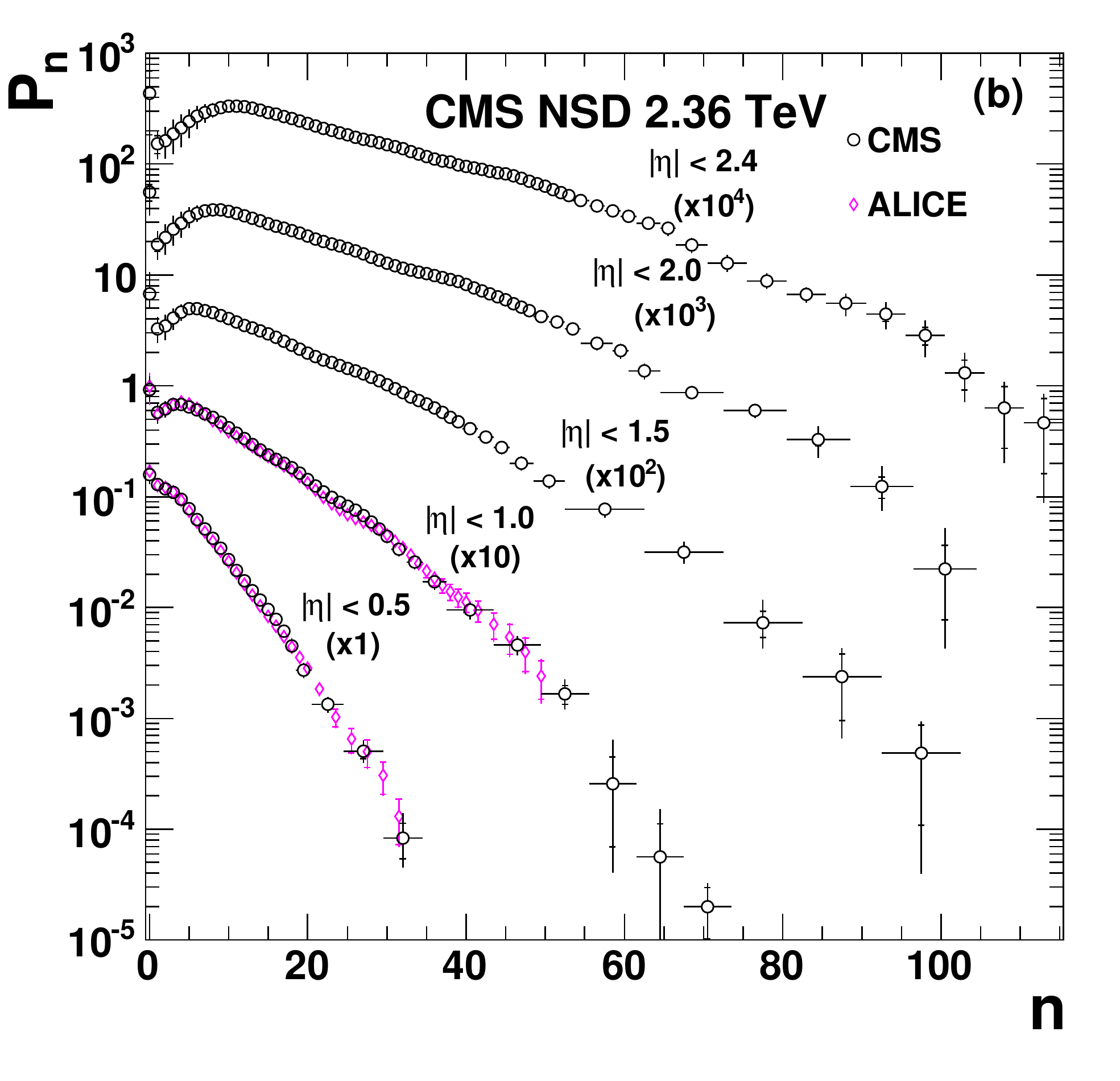}\\
\includegraphics[width=0.49\textwidth]{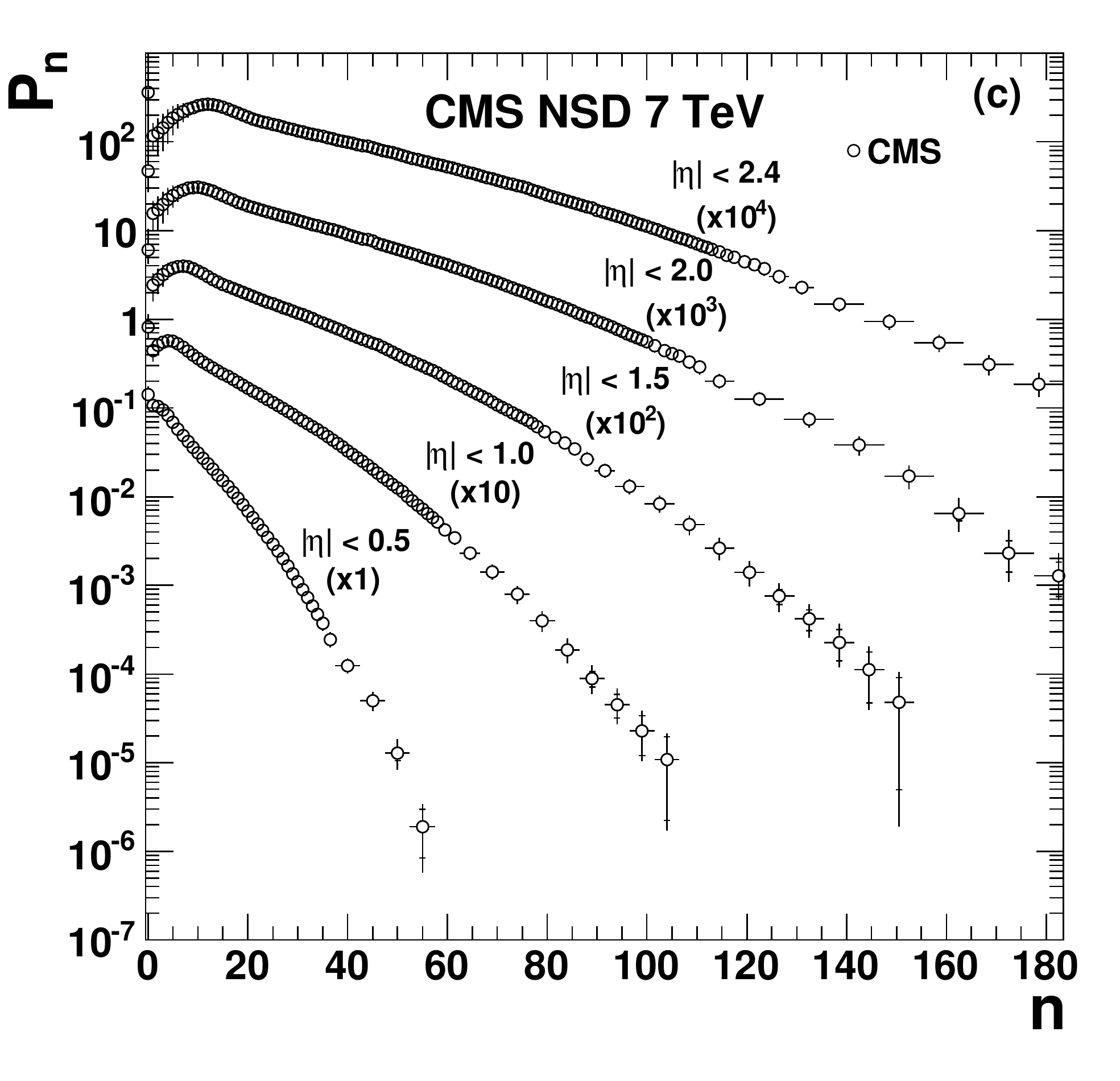} &
\end{tabular}
\caption{The fully corrected charged hadron multiplicity spectrum for $|\eta|<0.5$, 1.0, 1.5, 2.0, 
and 2.4, (a) at $\sqrt{s} = 0.9\TeV$, (b) 2.36\TeV, and (c) 7\TeV, compared with other measurements in the same $\eta$~interval
and at the same centre-of-mass energy~\cite{Alner:1984is,
Alner:1985wj,Aamodt:2010ft, Aamodt:2010pp}. For clarity, results in different
pseudorapidity intervals are scaled by powers of 10 as given in the plots.
The error bars are the statistical and systematic uncertainties added in quadrature. \label{UA5plot1}}
\end{center}
\end{figure}

The NSD charged hadron multiplicity distributions are measured in increasing ranges of
pseudorapidity from $|\eta|<0.5$ to $|\eta|<2.4$.
The fully corrected results at $\sqrt{s} = 0.9$, 2.36, and 7\TeV are
compared in Fig.~\ref{UA5plot1} with earlier measurements in the same
pseudorapidity ranges performed by the
UA5~\cite{Alner:1984is, Alner:1985wj} and ALICE~\cite{Aamodt:2010ft, Aamodt:2010pp} collaborations. 
Our measurements were also compared with results obtained with a
CMS cross-check analysis (not shown) on data at $\sqrt{s} = 0.9$ and 7\TeV,
using a tracklet-based tracking algorithm as in Ref.~\cite{Khachatryan:2010xs}. With a reconstruction efficiency
exceeding $90\%$ for $\pt>50\MeVc$, the latter
provided a cross-check of the extrapolation for tracks below $\pt < 100\MeVc$, including the use of data without magnetic
field at 7\TeV.
All measurements agree well within their total uncertainties.

In the largest pseudorapidity interval of $|\eta| < 2.4$, there is a
change of slope in $P_n$ for $n>20$, indicating a multicomponent structure as was
discussed in~\cite{Alexopoulos:1998bi} and~\cite{Giovannini:1997ce} in terms
of multiple-soft-Pomeron exchanges. 
This feature becomes more pronounced with increasing centre-of-mass energies, notably at $\sqrt{s} = 7\TeV$.

An extensive range of tunes~\cite{Bartalini:2008zz, Moraes:2007rq,
Buckley:2009bj, Skands:2010ak} based on the \PYTHIAV fragmentation
model have been developed. They differ mainly in their parametrisation of the multiple-parton-interaction model. 
Some reproduce the charged hadron multiplicities better than others, but
none is able to give a good description simultaneously at all the
centre-of-mass energies and in all pseudorapidity ranges. 
For clarity, only the baseline tune D6T is shown in comparison with other
models having a different physical description of soft-particle production such as
\PHOJET~\cite{Engel:1994vs, Engel:1995yda} and the new fragmentation model of \PYTHIAeight~\cite{Sjostrand:2007gs}.

\begin{figure}[hbt!]
\begin{center}
\begin{tabular}{cc}
\includegraphics[width=0.49\textwidth]{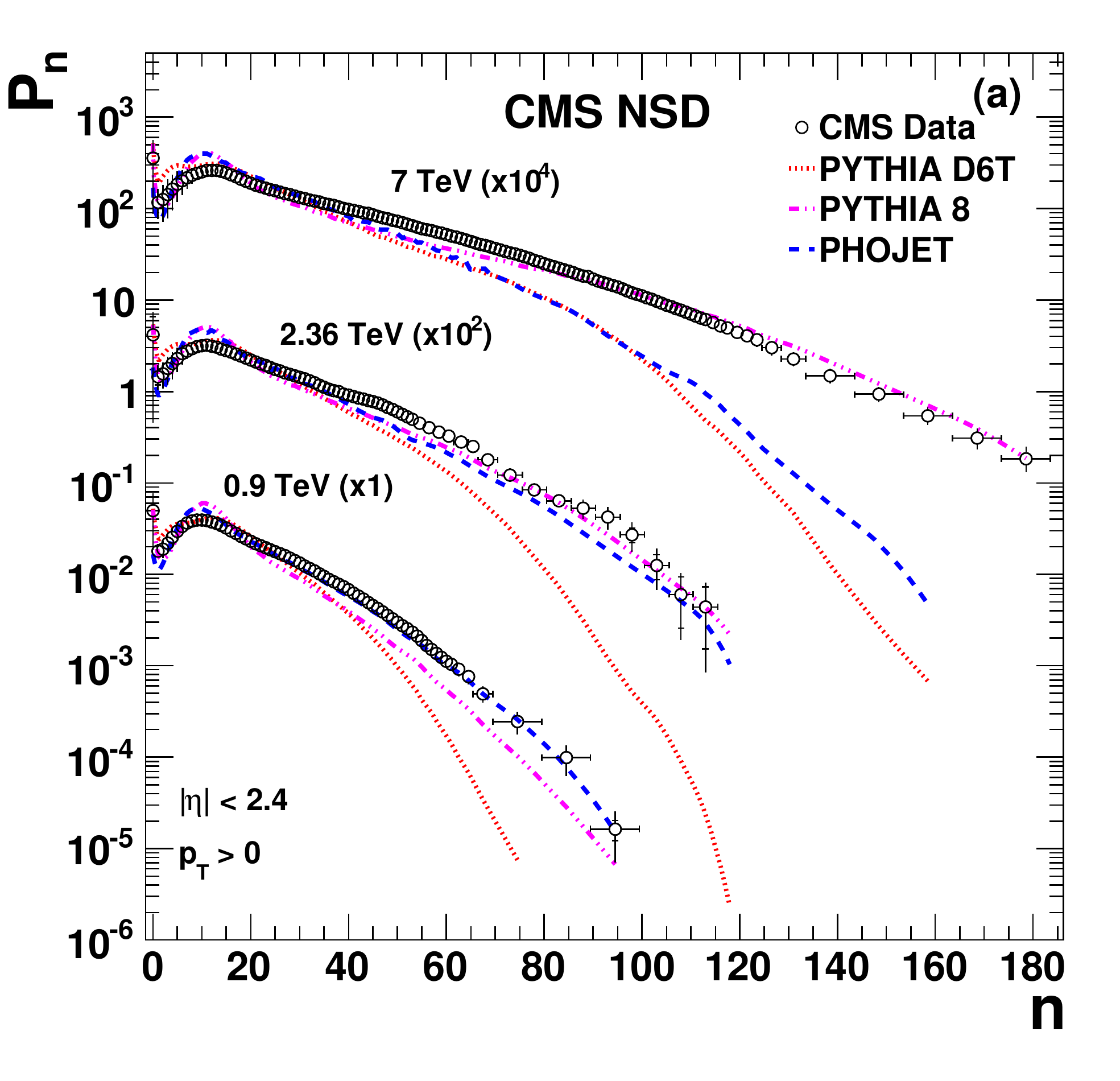} &
\includegraphics[width=0.49\textwidth]{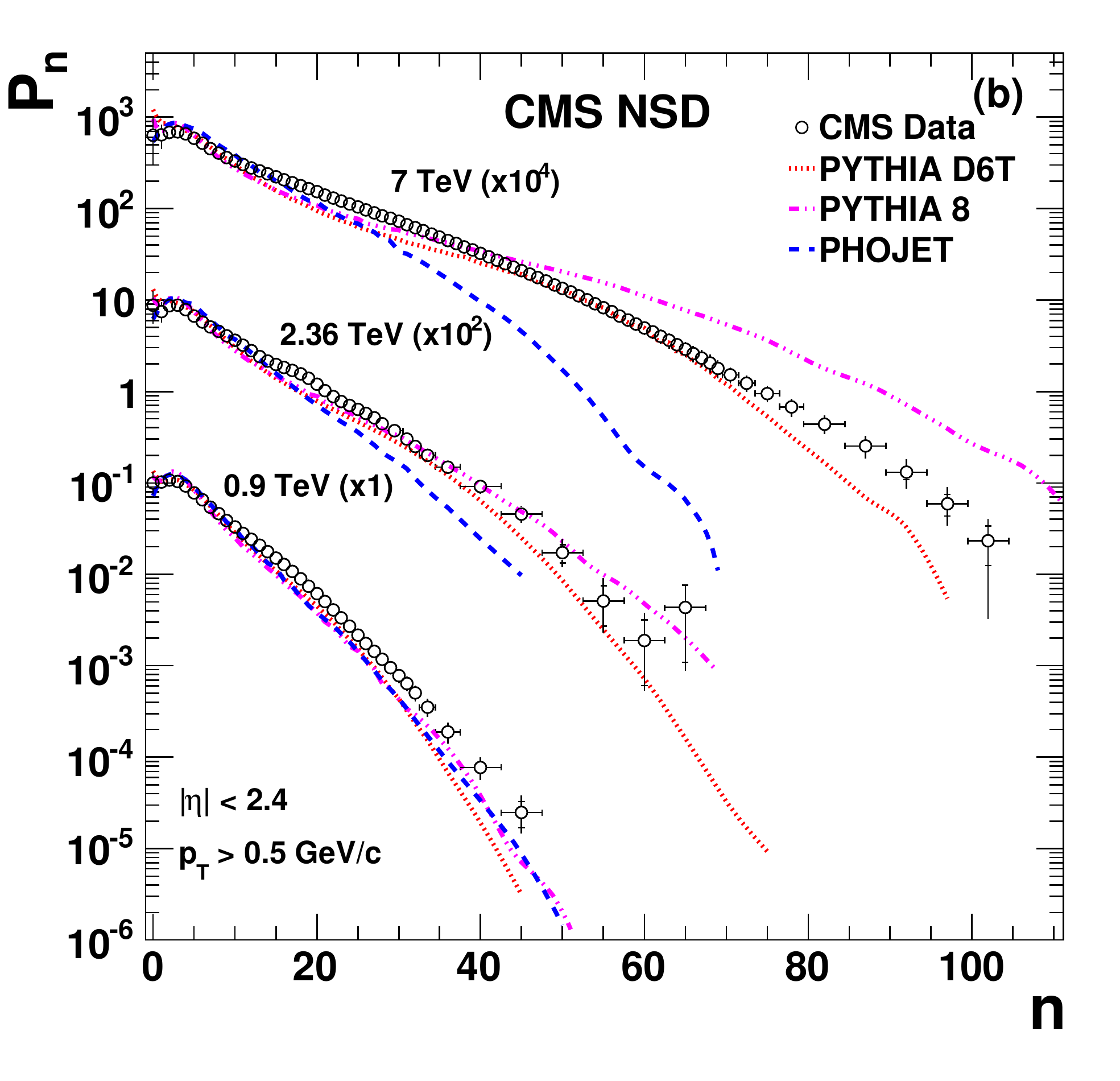}
\end{tabular}
\caption{The charged hadron multiplicity distributions with
 $|\eta|<2.4$ for (a) $\pt > 0$ and (b) $\pt > 500\MeVc$ at $\sqrt{s} = 0.9$, 2.36, and 7\TeV,
 compared to two different \PYTHIA models and the \PHOJET model. For clarity, results for different centre-of-mass
 energies are scaled by powers of 10 as given in the plots.\label{tunecompar}}
\end{center}
\end{figure}

\begin{figure}[hbt!]
\begin{center}
\includegraphics[width=0.55\textwidth]{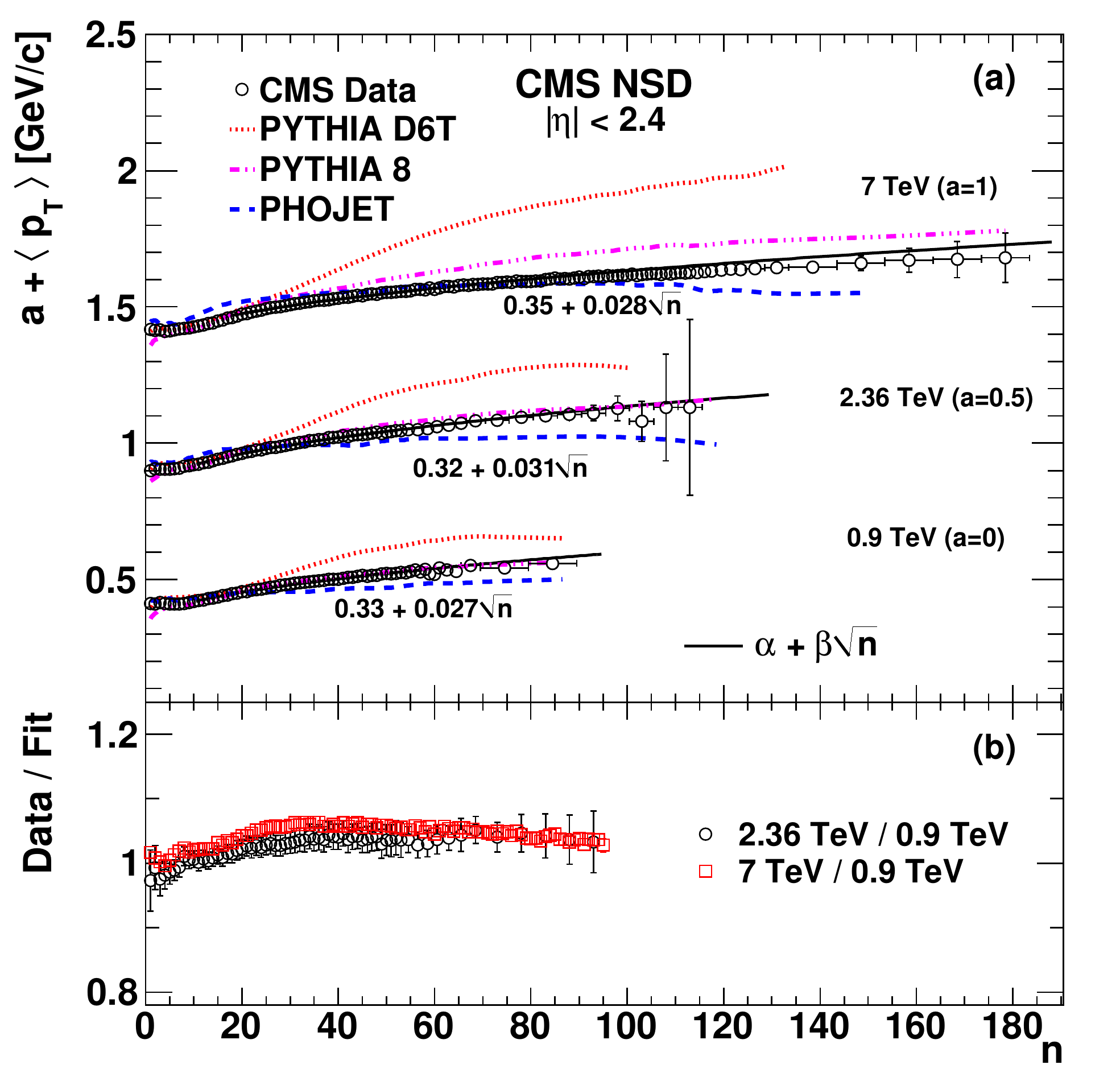}
\caption{(a) A comparison of $\langle \pt \rangle $ versus $n$ for
$|\eta|<2.4$ with two different \PYTHIA models and the \PHOJET model at
$\sqrt{s} = 0.9$, 2.36, and 7\TeV. For clarity, results for different energies are
shifted by the values of $a$ shown in the plots. Fits to the high-multiplicity part ($n>15$)
with a linear form in $\sqrt{n}$ are superimposed. (b) The ratios of the
higher-energy data to the fit at 0.9~TeV indicate the approximate
energy independence of $\langle \pt \rangle $ at fixed $n$. 
\label{ptvsn1}}
\end{center}
\end{figure}

A comparison of our measurements with these three classes of models is shown in Fig.~\ref{tunecompar}
for all charged hadrons and for those with $\pt > 500\MeVc$. \PYTHIADsixT
underestimates drastically the multiplicity at all measured energies but
improves when $\pt > 500\MeVc$ is required. \PYTHIAeight is the only model that gives a reasonable
description of the multiplicity distribution at all energies, but tends to
overestimate the multiplicity at 7~\TeV when $\pt > 500\MeVc$ is
required. \PHOJET produces too few charged hadrons overall but gives
a good description of the average transverse momentum $\langle \pt
\rangle$ at fixed multiplicity $n$, as illustrated in Fig.~\ref{ptvsn1}.  
Among the three classes of models, \PYTHIAeight gives the best overall description of the multiplicity
distribution and the dependence of the average transverse momentum on $n$. 
Inspired by~\cite{McLerran:2010ex} we fit a first-degree polynomial in 
$\sqrt{n}$ to the multiplicity dependence of $\langle \pt \rangle (n)$ for $n>15$ at each energy, 
yielding a good description which is valid at all three energies (Fig.~\ref{ptvsn1}).
The ratios of the data obtained
at 7~and~2.36~\TeV with respect to the data at 0.9~\TeV
show that the rise of the average transverse momentum with the multiplicity is roughly energy-independent .

All previous observations seem to
indicate that the Monte Carlo models produce too few particles
with low transverse momenta, especially at 7\TeV. The \PYTHIA models tend
to compensate for this by producing too many particles with high transverse
momentum, which is related to the modelling of semi-hard
multiple-parton interactions.

\subsection{Violation of KNO scaling}
\begin{figure}[hbt!]
\begin{center}
\begin{tabular}{cc}
\includegraphics[width=0.49\textwidth]{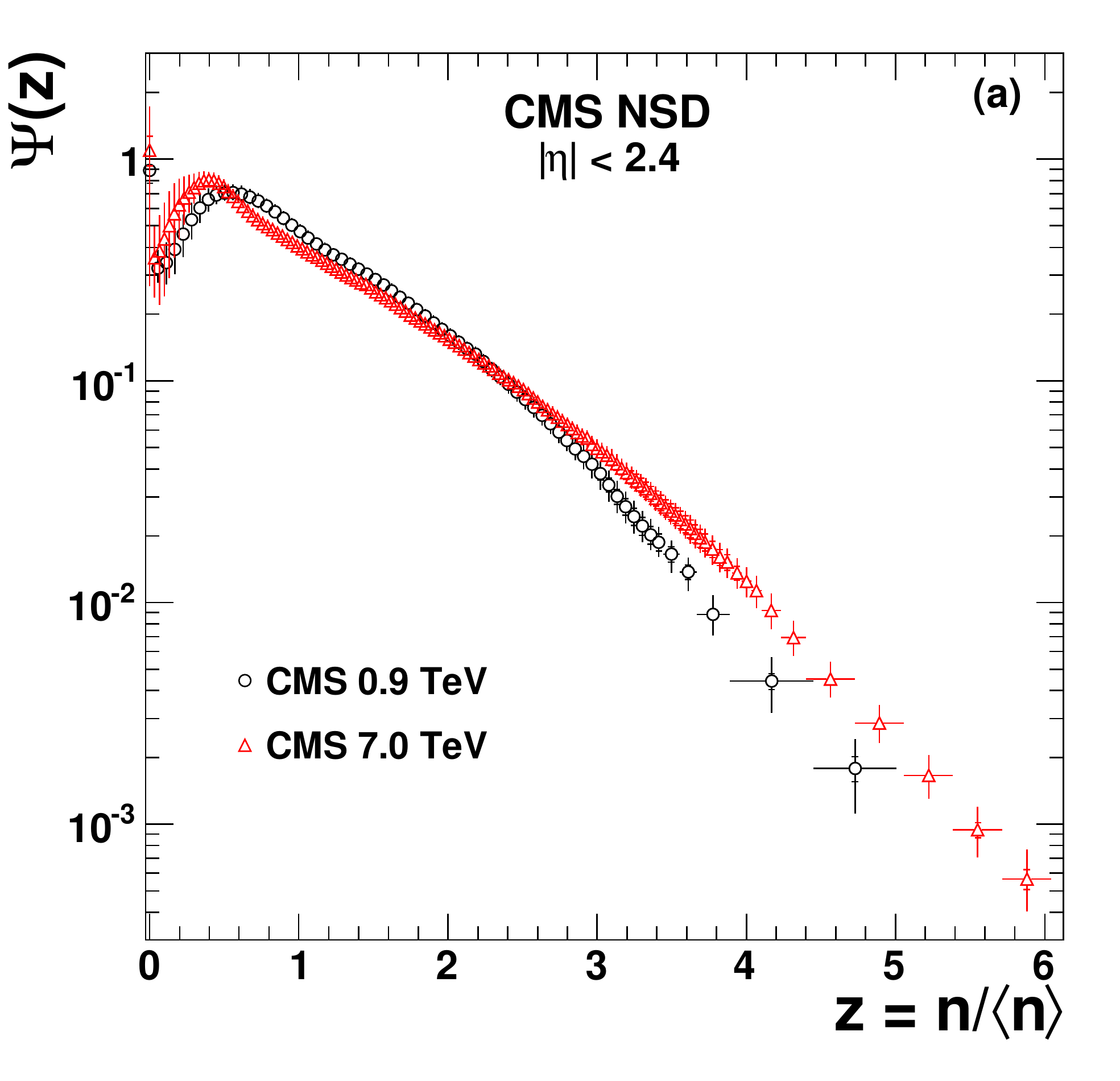} &
\includegraphics[width=0.49\textwidth]{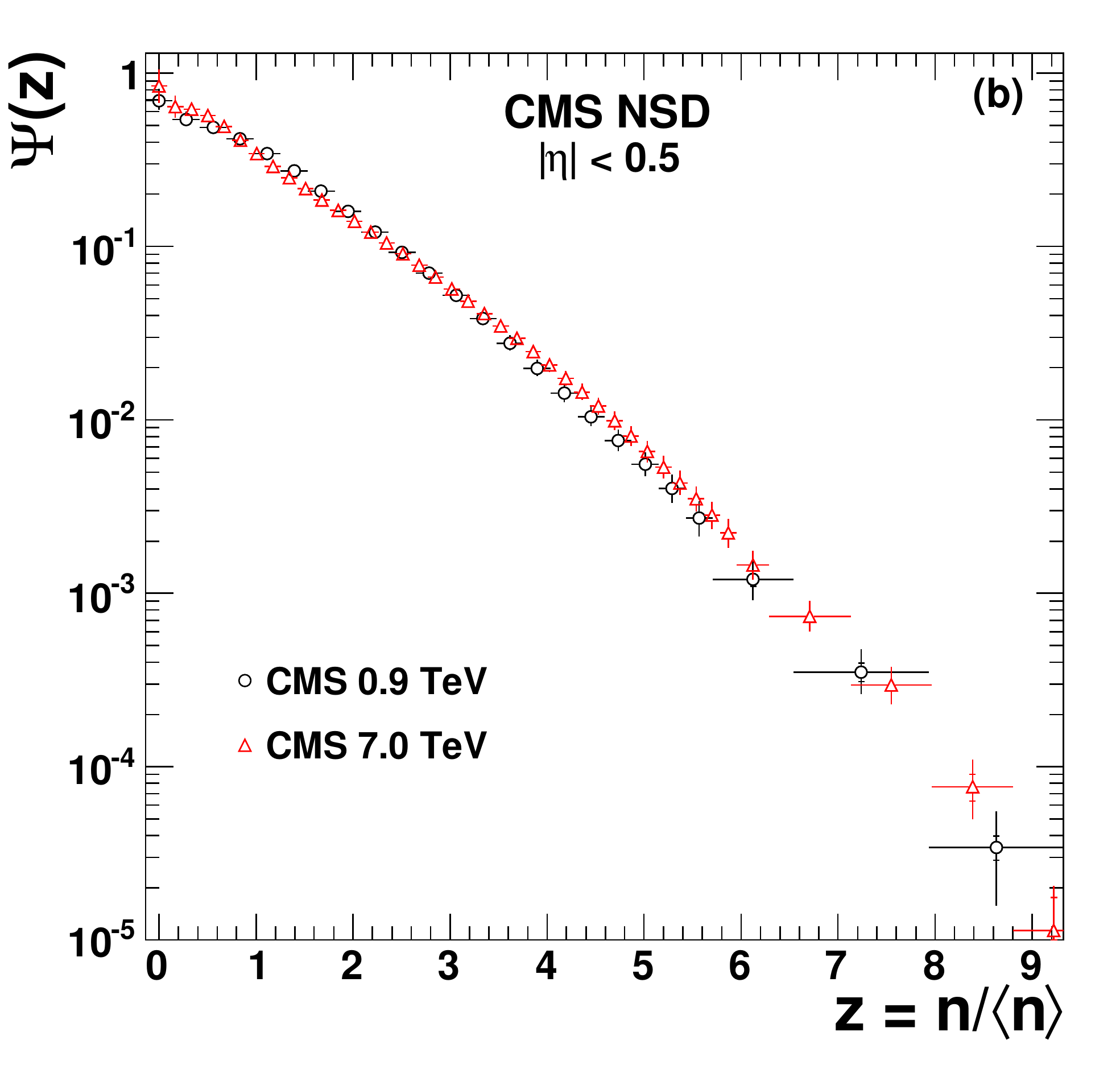}
\end{tabular}
\caption{The charged hadron multiplicity distributions in KNO form
at $\sqrt{s} = 0.9$ and 7\TeV in two pseudorapidity intervals, (a) $|\eta|<2.4$
and (b) $|\eta|<0.5$.\label{knoplots1} }
\end{center}
\end{figure}

The multiplicity distributions are shown in KNO form in Fig.~\ref{knoplots1}
for a large pseudorapidity interval of $|\eta|<2.4$, where we observe a strong
violation of KNO scaling between $\sqrt{s} = 0.9\TeV$ and 7\TeV, and for a small pseudorapidity
interval of $|\eta|<0.5$, where KNO scaling holds. 
Scaling is a characteristic property of the multiplicity
distribution in cascade processes of a single jet with self-similar
branchings and fixed coupling constant~\cite{Polyakov1, Polyakov2,Orfanidis:1974bd, CohenOchs, Dokshitzer:1993dc, Carius:1990cz, SzwedWrochna, Gazdzicki:1990bp}.

\begin{figure}[hbt!]
\begin{center}
\begin{tabular}{cc}
\includegraphics[width=0.49\textwidth]{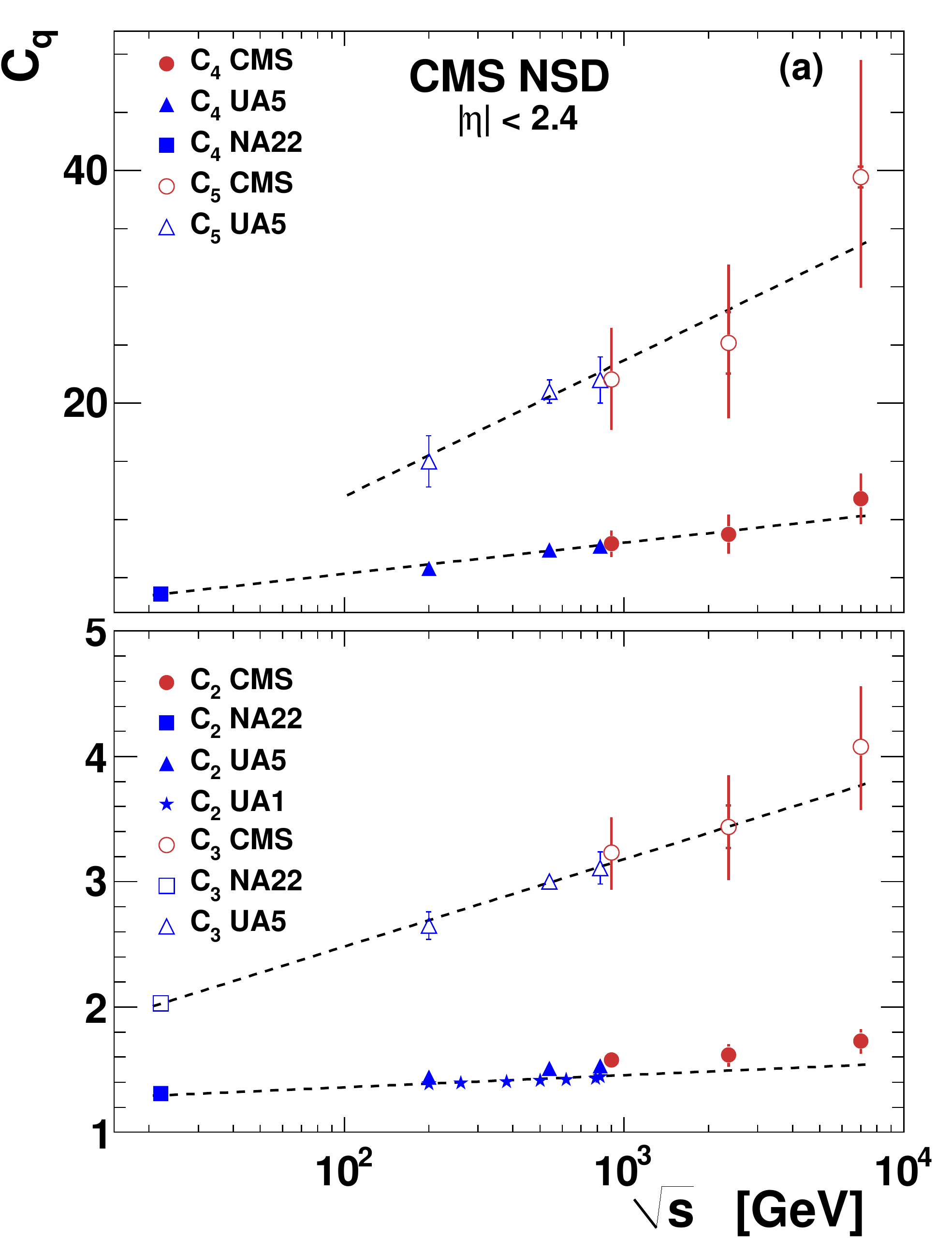} &
\includegraphics[width=0.49\textwidth]{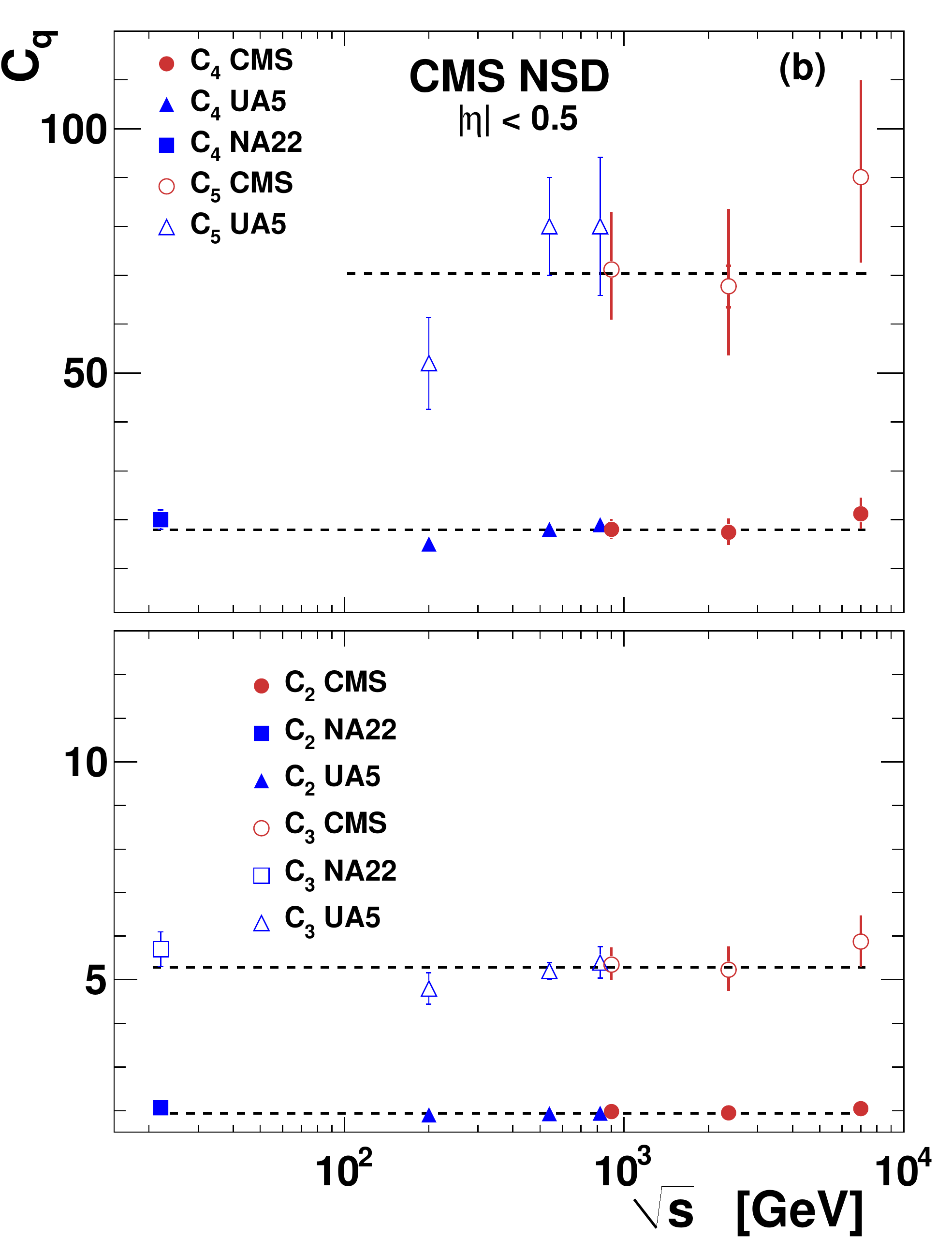}
\end{tabular}
\caption{Fits of the $\log s$ dependence of the normalised moments $C_q$ of the
multiplicity distribution for (a) $|\eta|<2.4$ (assuming linear dependence) and (b) $|\eta|<0.5$ (assuming no dependence),
including data from lower energy experiments~\cite{Adamus:1987ea,Alner:1985zc,Ansorge:1988kn}. For $\sqrt{s} = 0.9\TeV$, data
from experiments other than CMS were drawn shifted to lower $\sqrt{s}$ for clarity.
\label{moment1}}
\end{center}
\end{figure}

The validity of KNO scaling is shown more quantitatively in Fig.~\ref{moment1}
by the normalised order-$q$ moments $C_q$ of the multiplicity distribution, complemented with measurements
at lower energies~\cite{Adamus:1987ea, Alner:1985zc,Ansorge:1988kn}. For $|\eta|<2.4$ the values of $C_q$ increase linearly with
$\log s$, while for $|\eta|<0.5$ they remain constant up to
$q=4$ over the full centre-of-mass energy range, as illustrated by the fits in Fig.~\ref{moment1}.

Multiplicity distributions for  e$^+$e$^-$ annihilations up to the highest LEP energies show clear evidence for multiplicity
scaling, both in small ranges ($\Delta \eta<0.5$), in single hemispheres, and in full phase-space. However, at LEP
energies, scaling is broken for intermediate-size ranges where, besides two-jet events, multi-jet events contribute most prominently
\cite{Abreu:1992gp,Abreu:1991yc,Decamp:1991uz,Buskulic:1995xz,Acton:1991aa}.

For hadron-hadron collisions, approximate KNO scaling holds up to ISR energies~\cite{Thome:1977ky,
Breakstone:1983ns}, but clear scaling violations become manifest above
$\sqrt{s}\approx 200$ GeV both for the
multiplicity distributions in full phase space and in central pseudorapidity
ranges~\cite{Adamus:1987ea,Adamus:1986kx,Alexopoulos:1998bi, Acosta:2001rm}. In $\mathrm{p\overline{p}}$ collisions,  and
for large rapidity ranges, the UA5 experiment was the first to observe a larger than expected high-multiplicity
tail and a change of slope~\cite{Ansorge:1988kn, Alexopoulos:1998bi}, which was interpreted as evidence for a multi-component structure of the final
states~\cite{Fuglesangm, Matinyan:1998ja, Giovannini:1997ce}.
Our observation of strong KNO scaling violations at $\sqrt{s} = 7\TeV$, as
well as a change of slope in $P_n$, confirm these earlier measurements.

All these observations, together with  the sizable growth with energy of the non-diffractive inelastic cross
section, point to the increasing importance of multiple hard, semi-hard, and
soft partonic subprocesses in high energy hadron-hadron inelastic
collisions~\cite{Abramovskii:1972vi, Capella:1978rg, Capella:1992yb,
PhysRevD.45.92, Alexopoulos:1998bi, Matinyan:1998ja}.

\subsection{Energy dependence of the mean multiplicity}
The mean multiplicity $\langle n\rangle $ is the first moment of the multiplicity
distribution and is equal to the integral of the corresponding
single-particle inclusive density in the $\eta$ interval considered. The mean multiplicity $\langle n\rangle $ is observed
to rise with increasing centre-of-mass energy in hadron-hadron
collisions~\cite{Adamus:1987ea, Thome:1977ky, Breakstone:1983ns,Alner:1984is,
Alner:1985wj, Arnison:1982rm, Albajar:1989an, Rimondi:1993wq}. 
The same behaviour is also observed
in e$^+$e$^-$ collisions, in deep-inelastic scattering~\cite{Aid:1996cb}, and in heavy ion
collisions~\cite{Kittel:2005fu}.

\begin{figure}[hbt!]
\begin{center}
\includegraphics[width=0.55\textwidth]{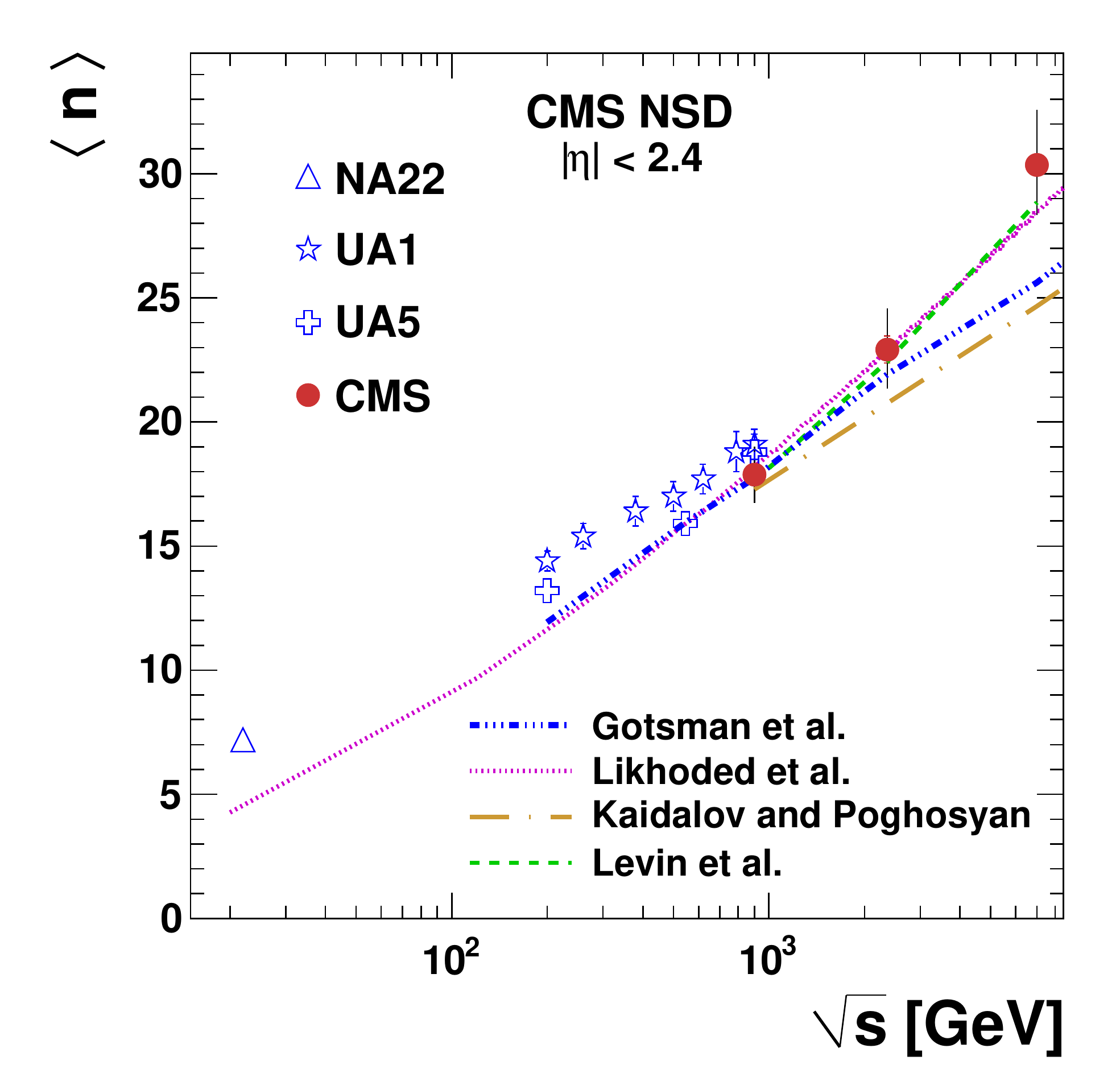}
\caption{The evolution of the mean charge multiplicity with the
centre-of-mass energy for $|\eta|<2.4$, including data from lower-energy experiments
for $|\eta|<2.5$~\cite{Adamus:1987ea,Alner:1985zc,Ansorge:1988kn,Albajar:1989an}.
The data are compared with predictions from three analytical Regge-inspired
models~\cite{Likhoded:2010pc,Gotsman:2009xm,Kaidalov:2009hn} and from a saturation model~\cite{Levin:2010dw}.\label{means1}}
\end{center}
\end{figure}

Our measured mean multiplicity is compared with experimental data obtained at
lower energies in Fig.~\ref{means1}.
Recent Regge-inspired models~\cite{Likhoded:2010pc,Gotsman:2009xm,Kaidalov:2009hn} predict a power-like behaviour among which
only Ref.~\cite{Likhoded:2010pc} describes the highest energy data very well. Parton saturation models (such
as~\cite{Levin:2010dw}) predict a strong rise of the central rapidity plateau as well.
Table~\ref{tabmnch} gives an overview of $\langle n\rangle $ for the data and for the
\PYTHIADsixT, \PYTHIAeight, and \PHOJET event generators.
As in Section~\ref{result1}, \PYTHIADsixT produces on average too few particles per event at all three energies. \PHOJET is
consistant with the data within uncertainties for $\sqrt{s} = 0.9\TeV$, but is not able to predict properly the mean
multiplicity at higher energies. \PYTHIAeight describes best the 7\TeV data, but underestimates $\langle n\rangle $
systematically at all energies.

\clearpage

\begin{table}[t!]
\begin{center}
\caption{Mean multiplicity for data, \PYTHIADsixT, \PYTHIAeight, and \PHOJET for $|\eta|<2.4$ at each centre-of-mass
energy. For data, the quoted uncertainties are first statistical, then upward and downward systematic.\label{tabmnch}}
\begin{tabular}{|c|c|c|c|c|}
\multicolumn{5}{c}{}\\
\hline
\multirow{2}{*}{$\sqrt{s}$~(TeV)} & \multicolumn{4}{c|}{$\langle n \rangle$} \\
\cline{2-5}
	& Data & \PYTHIADsixT & \PYTHIAeight & \PHOJET \\\hline
\multicolumn{1}{|c|}{0.9}     & $17.9 \pm 0.1^{+ 1.1}_{- 1.1}$ & 14.7 & 14.9 & 17.1 \\   
\multicolumn{1}{|c|}{2.36}    & $22.9 \pm 0.5^{+ 1.6}_{- 1.5}$ & 16.7 & 17.8 & 18.7 \\
\multicolumn{1}{|c|}{7}	      & $30.4 \pm 0.2^{+ 2.2}_{- 2.0}$ & 21.2 & 25.8 & 23.2 \\     
\hline
\end{tabular}
\end{center}
\end{table}

\section{Conclusions}
The charged hadron multiplicity distributions of non-single-diffractive
events were measured from an analysis of the minimum-bias datasets collected
by CMS at three centre-of-mass energies: $\sqrt{s} = 0 .9$, 2.36, and 7\TeV.
The excellent tracking capabilities of the silicon  pixel and strip detectors
of CMS, combined with an optimised tracking and vertexing algorithm, allow the
reconstruction of charged tracks down to $\pt = 100\MeVc$ with high efficiency
and low background contamination. A full correction for detector resolution
and acceptance effects and an extrapolation to zero transverse momentum yield measurements of
the charged hadron multiplicity distribution for
increasing central pseudorapidity ranges from $|\eta|<0.5$ to $|\eta|<2.4$, which can be compared
with models of soft-particle production and with experimental
data at lower energies.

Although some event generators provide an adequate description of
Tevatron and LEP data, none is able to describe simultaneously the multiplicity distributions
and the \pt spectrum at $\sqrt{s} = 7\TeV$. 
In general, models predict too few low-momentum particles, indicating that
by increasing the amount of multiple-parton interactions one effectively
introduces too many hard scatters in the event.

The change of slope in $P_n$ in the widest central pseudorapidity intervals observed at
$\sqrt{s} = 7\TeV$, combined with the strong linear increase of the $C_q$ moments,
indicates a clear violation of KNO scaling with respect to
lower energies. This observation merits further studies.

\section*{Acknowledgements}
\hyphenation{Ger-many}
\hyphenation{Bundes-ministerium Forschungs-gemeinschaft Forschungs-zentren Ger-many} We wish to congratulate our colleagues
in the CERN accelerator departments for the excellent performance of the LHC machine. We thank the technical and
administrative staff at CERN and other CMS institutes. This work was supported by the Austrian Federal Ministry of Science
and Research; the Belgium Fonds de la Recherche Scientifique, and Fonds voor Wetenschappelijk Onderzoek; the Brazilian
Funding Agencies (CNPq, CAPES, FAPERJ, and FAPESP); the Bulgarian Ministry of Education and Science; CERN; the Chinese
Academy of Sciences, Ministry of Science and Technology, and National Natural Science Foundation of China; the Colombian
Funding Agency (COLCIENCIAS); the Croatian Ministry of Science, Education and Sport; the Research Promotion Foundation,
Cyprus; the Estonian Academy of Sciences and NICPB; the Academy of Finland, Finnish Ministry of Education, and Helsinki
Institute of Physics; the Institut National de Physique Nucl\'eaire et de Physique des Particules~/~CNRS, and Commissariat
\`a l'\'Energie Atomique, France; the Bundesministerium f\"ur Bildung und Forschung, Deutsche Forschungsgemeinschaft, and
Helmholtz-Gemeinschaft Deutscher Forschungszentren, Germa-ny; the General Secretariat for Research and Technology, Greece; the National Scientific Research Foundation, and National Office for Research and Technology, Hungary; the Department of Atomic Energy, and Department of Science and Technology, India; the Institute for Studies in Theoretical Physics and Mathematics, Iran; the Science Foundation, Ireland; the Istituto Nazionale di Fisica Nucleare, Italy; the Korean Ministry of Education, Science and Technology and the World Class University program of NRF, Korea; the Lithuanian Academy of Sciences; the Mexican Funding Agencies (CINVESTAV, CONACYT, SEP, and UASLP-FAI); the Pakistan Atomic Energy Commission; the State Commission for Scientific Research, Poland; the Funda\c{c}\~ao para a Ci\^encia e a Tecnologia, Portugal; JINR (Armenia, Belarus, Georgia, Ukraine, Uzbekistan); the Ministry of Science and Technologies of the Russian Federation, and Russian Ministry of Atomic Energy; the Ministry of Science and Technological Development of Serbia; the Ministerio de Ciencia e Innovaci\'on, and Programa Consolider-Ingenio 2010, Spain; the Swiss Funding Agencies (ETH Board, ETH Zurich, PSI, SNF, UniZH, Canton Zurich, and SER); the National Science Council, Taipei; the Scientific and Technical Research Council of Turkey, and Turkish Atomic Energy Authority; the Science and Technology Facilities Council, UK; the US Department of Energy, and the US National Science Foundation.

Individuals have received support from the Marie-Curie programme and the European Research Council (European Union); the Leventis Foundation; the A. P. Sloan Foundation; the Alexander von Humboldt Foundation; the Associazione per lo Sviluppo Scientifico e Tecnologico del Piemonte (Italy); the Belgian Federal Science Policy Office; the Fonds pour la Formation \`a la Recherche dans l'\'industrie et dans l'\'Agriculture (FRIA-Belgium); and the Agentschap voor Innovatie door Wetenschap en Technologie (IWT-Belgium).

\newpage
\clearpage
\bibliography{auto_generated}

\cleardoublepage\appendix\section{The CMS Collaboration \label{app:collab}}\begin{sloppypar}\hyphenpenalty=5000\widowpenalty=500\clubpenalty=5000\textbf{Yerevan Physics Institute,  Yerevan,  Armenia}\\*[0pt]
V.~Khachatryan, A.M.~Sirunyan, A.~Tumasyan
\vskip\cmsinstskip
\textbf{Institut f\"{u}r Hochenergiephysik der OeAW,  Wien,  Austria}\\*[0pt]
W.~Adam, T.~Bergauer, M.~Dragicevic, J.~Er\"{o}, C.~Fabjan, M.~Friedl, R.~Fr\"{u}hwirth, V.M.~Ghete, J.~Hammer\cmsAuthorMark{1}, S.~H\"{a}nsel, C.~Hartl, M.~Hoch, N.~H\"{o}rmann, J.~Hrubec, M.~Jeitler, G.~Kasieczka, W.~Kiesenhofer, M.~Krammer, D.~Liko, I.~Mikulec, M.~Pernicka, H.~Rohringer, R.~Sch\"{o}fbeck, J.~Strauss, A.~Taurok, F.~Teischinger, W.~Waltenberger, G.~Walzel, E.~Widl, C.-E.~Wulz
% ...
\vskip\cmsinstskip
% ...
1:~~Also at CERN, European Organization for Nuclear Research, Geneva, Switzerland\\
% ...

\end{sloppypar}
\end{document}